%% file: BPH-21-006_temp.tex
\pdfoutput=1
\documentclass[11pt,twoside,a4paper,cmspaper,final,collab]{cms-tdr}
\def\svnVersion{69a14e7}\def\svnDate{2023/05/23}\def\cmsCernNoTag{CERN-EP-2022-270}\def\cmsCernDate{\today}\def\cmsMessage{Published in Physics Letters B as \href{http://dx.doi.org/10.1016/j.physletb.2023.137955}{\doi{10.1016/j.physletb.2023.137955}.}}
\begin{document}\cmsNoteHeader{BPH-21-006}

\newcommand{\cbf}       {\ensuremath{\mathcal{B}}}
\newcommand{\eps}       {\varepsilon\xspace}
\newcommand{\Bd}        {\PBz}
\newcommand{\Bs}        {\PBzs}
\newcommand{\fu}        {\ensuremath{f_\PQu}\xspace}
\newcommand{\fd}        {\ensuremath{f_\PQd}\xspace}
\newcommand{\fs}        {\ensuremath{f_\PQs}\xspace}
\newcommand{\fsfu}      {\ensuremath{\fs/\fu}\xspace}
\newcommand{\fufd}      {\ensuremath{\fu/\fd}\xspace}
\newcommand{\BsH}       {\ensuremath{{\PB}^0_{\PQs,\mathrm{H}}}\xspace}
\newcommand{\BsL}       {\ensuremath{{\PB}^0_{\PQs,\mathrm{L}}}\xspace}
\newcommand{\mvab}      {\ensuremath{d_{\text{MVA}}}\xspace}
\newcommand{\bsll}      {\ensuremath{\PQb \to \PQs\ell^+\ell^-}\xspace}
\newcommand{\bkstmm}    {\ensuremath{\PB\to \PKst\MM}\xspace}
\newcommand{\bdmm}      {\ensuremath{\Bd \to \MM}\xspace}
\newcommand{\bsmm}      {\ensuremath{\Bs \to\MM}\xspace}
\newcommand{\bjpsik}    {\ensuremath{\PBp \to \PJGy\PKp}\xspace}
\newcommand{\bjpsiphi}  {\ensuremath{\Bs\to\PJGy\PGfP{1020}}\xspace}
\newcommand{\bhh}       {\ensuremath{\PB\to  \Ph^+\Ph^-}\xspace}
\newcommand{\PJGymm}    {\ensuremath{\PJGy\to\MM}\xspace}
\newcommand{\bmm}       {\ensuremath{\PB\to\MM}\xspace}
\newcommand{\kspipi}    {\ensuremath{\PKzS\to\PGpp\PGpm}\xspace}
\newcommand{\phikk}     {\ensuremath{\PGfP{1020} \to \PKp\PKm}\xspace}
\newcommand{\lppi}      {\ensuremath{\PGL \to \Pp\PGpm}\xspace}
\newcommand{\ptmu}      {\ensuremath{\pt(\PGm)}\xspace}
\newcommand{\etamu}     {\ensuremath{\eta(\PGm)}\xspace}
\newlength\cmsTabSkip\setlength{\cmsTabSkip}{1ex}
\ifthenelse{\boolean{cms@external}}{\providecommand{\NA}{\ensuremath{\cdots}\xspace}}{\providecommand{\NA}{\ensuremath{\text{---}}\xspace}}
\ifthenelse{\boolean{cms@external}}{\providecommand{\cmsLeft}{upper\xspace}}{\providecommand{\cmsLeft}{left\xspace}}
\ifthenelse{\boolean{cms@external}}{\providecommand{\cmsRight}{lower\xspace}}{\providecommand{\cmsRight}{right\xspace}}
\ifthenelse{\boolean{cms@external}}{\providecommand{\cmsTopRight}{middle\xspace}}{\providecommand{\cmsTopRight}{right\xspace}}
\ifthenelse{\boolean{cms@external}}{\providecommand{\cmsTwoUpper}{top and middle plots\xspace}}{\providecommand{\cmsTwoUpper}{upper plots\xspace}}
\ifthenelse{\boolean{cms@external}}{\providecommand{\largePlotWidth}{0.48\textwidth}}{\providecommand{\largePlotWidth}{0.65\textwidth}}
\providecommand{\cmsTable}[1]{\resizebox{\textwidth}{!}{#1}}
\ifthenelse{\boolean{cms@external}}{\providecommand{\tableCMS}[1]{\resizebox{\columnwidth}{!}{#1}}}{\providecommand{\tableCMS}[1]{#1}}

\cmsNoteHeader{BPH-21-006}

\title{Measurement of the \texorpdfstring{\bsmm}{B0s to mu+ mu-} decay properties and search for the \texorpdfstring{\bdmm}{B0 to mu+ mu-} decay in proton-proton collisions at \texorpdfstring{$\sqrt{s}=13\TeV$}{sqrt(s) = 13 TeV}}

\author*[cern]{The CMS Collaboration}

\date{\today}

\abstract{Measurements are presented of the \bsmm branching fraction and effective lifetime, as well as results of a search for the \bdmm decay in proton-proton collisions at $\sqrt{s}=13\TeV$ at the LHC. The analysis is based on data collected with the CMS detector in 2016--2018 corresponding to an integrated luminosity of 140\fbinv. The branching fraction of the \bsmm decay and the effective \Bs meson lifetime are the most precise single measurements to date. No evidence for the \bdmm decay has been found. All results are found to be consistent with the standard model predictions and previous measurements.}

\hypersetup{
pdfauthor={CMS Collaboration},%
pdftitle={Measurement of the B0s to mu+ mu- decay properties and search for the B0 to mu+ mu- decay in proton-proton collisions at sqrt(s) = 13 TeV},%
pdfsubject={CMS},
pdfkeywords={CMS, BPH, flavour physics, rare decays}}

\maketitle 

\section{Introduction}
\label{sec:introduction}

Rare \PB meson decays provide a sensitive probe for beyond-the-standard-model
(BSM) effects and allow exploring energy scales much higher than the ones directly
accessible at the CERN LHC. A key factor in the
success of these studies is the availability of precise theoretical predictions
for experimentally accessible processes. The leptonic decays \bsmm and \bdmm represent such a 
case, where precise theoretical predictions can be matched with a clear
experimental signature. 
These rare decays are examples of flavor changing neutral current processes,
which are strongly suppressed in the standard model (SM), making them sensitive
to BSM physics contributions. In this Letter, when decays are mentioned,
charge-conjugated decay modes are implied.

The \bsmm and \bdmm decays proceed through penguin and box diagrams that involve \PZ or \PW\ boson exchange and are
furthermore helicity suppressed by a factor $m_\PGm^2/m_{\PB}^2$, where $m_\PGm$ and
$m_{\PB}$ denote the masses of the muon and either the \Bs or \Bd meson. 
Moreover, the \bdmm decay is also Cabibbo--Kobayashi--Maskawa suppressed~\cite{PhysRevLett.10.531,PTP.10.1143}.
As a result, in the SM, the average
time-integrated branching fractions for these decays are very
small~\cite{Beneke:2019slt}:
 \begin{linenomath}
 \begin{align*}
 \cbf(\bsmm) &= (3.66\pm0.14)\times 10^{-9},\\
 \cbf(\bdmm) &= (1.03\pm0.05)\times 10^{-10}.
\end{align*}
 \end{linenomath}
These predictions include next-to-leading order corrections of electroweak
origin and next-to-next-to-leading order quantum chromodynamics (QCD) corrections. The largest
contribution to the theoretical uncertainty is from the determining of the
Cabibbo--Kobayashi--Maskawa matrix element values, in particular $\vert V_{\PQc\PQb}\vert$.
The issue can be mitigated by considering a ratio of the branching fraction to the mass difference
of the heavy and light \Bs and \Bd mesons~\cite{Bobeth:2021cxm}.

A number of experiments at $\EE$ and hadron colliders have searched for these decays,
but only recently the first observation of the \bsmm decay was reported in a
combined analysis of data taken by the CMS and LHCb
Collaborations~\cite{CMS:2014xfa}, which was later confirmed by the
ATLAS~\cite{Aaboud:2018mst}, CMS~\cite{CMS:2019bbr}, and LHCb~\cite{LHCb:2017rmj,LHCb:2021awg,LHCb:2021awg2}
experiments individually. Currently, the most
precise measurement of the \bsmm branching fraction is achieved in a combined analysis of data from the three
experiments~\cite{CMS:2020rox}. The analysis shows a deviation from the SM prediction at the level of 2.4 standard deviations ($\sigma$) for the \bsmm branching fraction. 
No significant detection of the \bdmm decay has been reported so far.

A few recent measurements of the semileptonic \bsll processes (where lepton
$\ell = \Pe$ or \Pgm) have reported disagreements at the level of
2--3\unit{$\sigma$} with the SM predictions.  Deviations were found in
measurements of the differential branching fractions of the \bkstmm and $\Bs
\to\PGf\MM$ decays~\cite{LHCb:2014cxe,LHCb:2021wdu,LHCb:2016ykl,LHCb:2021zwz},
angular observables in the \bkstmm decays~\cite{LHCb:2020lmf,LHCb:2020gog}, and
in searches for lepton flavor universality violation via measurements of the
$R_\PK$ and $R_\PKst$
ratios~\cite{LHCb:2017avl,LHCb:2021trn,Belle:2019oag,BELLE:2019xld,LHCb:2021lvy}. Not
all the individual measurements confirm these
observations though~\cite{CMS:2015bcy,CMS:2017rzx} and the revised $R_\PK$ and
$R_\PKst$ measurements are found to be consistent with the
SM~\cite{LHCb:2022qnv,LHCb:2022zom}.

In the framework of an effective field theory, the \bsll decays are dominated by the
semileptonic operators
$O_{9}=(\PAQs_L\gamma_\mu\PQb_L)(\overline{\ell}\gamma^\mu\ell)$ and
$O_{10}=(\PAQs_L\gamma_\mu\PQb_L)(\overline{\ell}\gamma^\mu\gamma_5\ell)$.
The BSM physics contributions could be observed as deviations of the
corresponding Wilson coefficients ($C_9$ and $C_{10}$) from their SM values.
As the \bsmm and \bdmm decays are dominated by the $O_{10}$ operator, they may be
sensitive to the same effects.
In contrast to the semileptonic case, the nonperturbative hadronic contributions
for leptonic \PB meson decays enter solely through decay constants $f_{\PB}$ and $f_{\PB_S}$, which are
precisely known from lattice QCD calculations~\cite{CMS:latticeQCD}, making the
theory calculations more robust. Therefore, a precise measurement of the \bsmm
decay properties may have a big impact on the interpretation of these anomalies.

The effective lifetime of the \Bs meson measured in the \bsmm decay is an
independent and theoretically clean probe for BSM physics~\cite{HFLAV:2022pwe}.
In the SM, the heavy (\BsH) and light (\BsL) mass eigenstates are linear
combinations of the flavor eigenstates. The lifetimes of the heavy and light mass
eigenstates are $\tau_{\mathrm{H}}= 1.624\pm0.009\unit{ps}$ and $\tau_{\mathrm{L}}=
1.429\pm0.007\unit{ps}$, respectively~\cite{Workman:2022ynf}. Only the heavy
\BsH mass eigenstate can decay to the \MM final state. This is because, in the
absence of charge-conjugation and parity ($CP$) violation in the \Bs mixing, the \Bs mass eigenstates are also $CP$ eigenstates,
with the heavier one being $CP$ odd, while \MM is also a $CP$-odd final
state. Therefore, any significant deviation of the measurement from the
established value for the \BsH lifetime would indicate a BSM physics
contribution. Currently, the most precise measurement of the \Bs meson
lifetime ($\tau$) in the \bsmm decay, $\tau(\bsmm) = 2.07 \pm 0.29\unit{ps}$,
comes from the LHCb experiment~\cite{LHCb:2021awg,LHCb:2021awg2}.

In this Letter, we report on new measurements of the \bsmm decay properties and a search
for the \bdmm decay based on proton-proton collision data collected by the CMS
experiment in 2016--2018 at a center-of-mass energy of 13 TeV,
corresponding to an integrated luminosity of 140\fbinv. This new analysis
uses improved techniques compared to the earlier publication based on 2011--2012
and 2016 data~\cite{CMS:2019bbr}; consequently the 2016 data sample has been
re-analyzed, and the new results supersede the ones from
Ref.~\cite{CMS:2019bbr}. Given that the sensitivity of the 2016--2018 data
significantly exceeds that of the 2011--2012 sample, no attempt is made to
combine the new results with those from 2011--2012 data.
Tabulated results are provided in the HEPData record for this analysis~\cite{hepdata}.

\section{Data analysis overview}
\label{sec:analysis_strategy}

The data analysis strategy employed in this measurement is based on the previous
CMS studies of the \bsmm and \bdmm decays. We made several modifications and
improvements to increase the analysis sensitivity, while also benefiting from the large amount
of data collected in 2017--2018. Because of the increased sensitivity, new methods
were developed to achieve a better understanding of various systematic effects.

Leptonic \PB meson decays are reconstructed by combining two oppositely charged muons,
performing a common vertex fit, and imposing selection criteria to separate small
signals from large backgrounds. Here and in what follows, we use the notation \PB 
to denote either the \Bd or \Bs meson. The dominant background sources are
the combinatorial background where the two muons originate from two different
heavy quarks, the partially reconstructed semileptonic decays where both muons
originate from the same \PB meson, and the peaking background coming from the
charmless two-body hadronic decays of \PB mesons.

The combinatorial and partially reconstructed backgrounds are the main limiting
factors in the analysis sensitivity. Despite being reducible backgrounds with
several distinct features, they are copious, which makes it difficult to reject
them completely without losing a significant fraction of signal events. To
maximize the analysis sensitivity, we perform a multivariate analysis (MVA)
combining multiple discriminating observables in a single powerful discriminator (\mvab)
using a boosted decision tree algorithm. The training and performance evaluation
of the MVA are described in Section~\ref{sec:mva}.

The charmless two-body decays, such as $\Bd \to \PKp\PGpm$ and $\Bs \to
\PKp\PKm$, could mimic signal when both charged hadrons are misidentified
as muons. We measure the misidentification probabilities in data using the \kspipi,
\phikk, and \lppi decays after restricting the decay distance to match that of the \PB mesons. 
We find a reasonable agreement between the data and
Monte Carlo (MC) simulation for pions and kaons, where the misidentification
probability is dominated by pion and kaon decays to a muon and a
muon antineutrino. The probability to misidentify protons as muons is an order of
magnitude smaller according to simulation, and found to be even smaller in data, thus
making contributions from the associated processes unimportant. With a stringent muon
identification based on a multivariate analysis~\cite{CMS:2019bbr}, we reduce the
charmless two-body backgrounds to a negligible level compared to the dominant combinatorial background.

The results are extracted using simultaneous unbinned maximum likelihood (UML)
fits in multiple categories (discussed in Section~\ref{sec:fitting}).
For the branching fraction measurements, we perform a two-dimensional (2D) fit using the dimuon
invariant mass ($m_{\MM}$) and its uncertainty as the observables. For the lifetime
extraction, we perform a three-dimensional (3D) fit using the dimuon mass, the decay time, and the decay time
uncertainty as the observables.

Given the poor precision in the knowledge of the \bbbar cross section at the LHC,
a direct extraction of the branching fractions of the \bsmm\ and \bdmm\ decays
would be affected by a large uncertainty. As commonly done in \PB physics
analyses, the signal branching fraction is instead calculated by normalizing it to
another decay channel of a \PB meson, for which the branching fraction is well known and
whose characteristics allow for a precise reconstruction with small backgrounds.
The \bjpsik decay with \PJGymm is the best candidate in our case as a
normalization channel. We also consider
\bjpsiphi decays with \phikk as a cross-check of the nominal result. The signal branching
fractions $\cbf(\bsmm)$ and $\cbf(\bdmm)$ can then be extracted as
\begin{linenomath} 
\ifthenelse{\boolean{cms@external}}
{ 
\begin{multline}
\cbf(\bsmm) = \cbf(\bjpsik)\frac{N_{\bsmm}}{N_{\bjpsik}} \\  \times\frac{\eps_{\bjpsik}}{\eps_{\bsmm}} \frac{\fu}{\fs}, \label{eq:bf_bs_jpsik}
\end{multline}
\begin{multline}
\cbf(\bsmm) = \cbf(\bjpsiphi)\frac{N_{\bsmm}}{N_{\bjpsiphi}} \\  \times\frac{\eps_{\bjpsiphi}}{\eps_{\bsmm}} \label{eq:bf_bs_jpsiphi},
\end{multline}
\begin{multline}
\cbf(\bdmm) = \cbf(\bjpsik)\frac{N_{\bdmm}}{N_{\bjpsik}} \\  \times\frac{\eps_{\bjpsik}}{\eps_{\bdmm}} \frac{\fu}{\fd}, \label{eq:bf_bd}
\end{multline}
} 
{ 
\begin{align}
\cbf(\bsmm) &= \cbf(\bjpsik)\frac{N_{\bsmm}}{N_{\bjpsik}} \frac{\eps_{\bjpsik}}{\eps_{\bsmm}} \frac{\fu}{\fs}, \label{eq:bf_bs_jpsik}
\\[1em]
\cbf(\bsmm) &= \cbf(\bjpsiphi)\frac{N_{\bsmm}}{N_{\bjpsiphi}} \frac{\eps_{\bjpsiphi}}{\eps_{\bsmm}} \label{eq:bf_bs_jpsiphi},
\\[1em]
\cbf(\bdmm) &= \cbf(\bjpsik)\frac{N_{\bdmm}}{N_{\bjpsik}} \frac{\eps_{\bjpsik}}{\eps_{\bdmm}} \frac{\fu}{\fd}, \label{eq:bf_bd}
\end{align}
} 
\end{linenomath} 
where $N_\PX$ is the number of the candidates of decay \PX, as extracted from the
fit, and $\eps_\PX$ is the corresponding full selection efficiency derived from MC simulation. In addition,
\fu, \fd, and \fs are the production fractions
for \PBp, \Bd, and \Bs mesons, respectively. The ratio \fufd is expected to be
1 in the SM due to isospin symmetry. The ratio \fsfu, as well as
$\cbf(\bjpsik)$ and $\cbf(\bjpsiphi)$, are external inputs discussed in
Section~\ref{sec:results}.

Another advantage of measuring the branching fractions with respect to other
decays is that it allows for a cancellation of many systematic uncertainties in
the selection and reconstruction efficiencies of the signal and normalization channels.

To reduce unintentional bias, the analysis employs a ``data blinding"
technique. All optimization studies are performed using signal from MC
simulation and background from data that does not include events with a dimuon
mass of 5.15--5.50\GeV.
Once the selection criteria and measurement procedure have been finalized, the
data are unblinded.

\section{The CMS detector}
\label{sec:detector}

The central feature of the CMS apparatus is a superconducting solenoid of
6\unit{m} internal diameter, providing a magnetic field of 3.8\unit{T}. Within
the solenoid volume are a silicon pixel and strip tracker, a lead tungstate
crystal electromagnetic calorimeter, and a brass and scintillator hadron
calorimeter, each composed of a barrel and two endcap sections. Forward
calorimeters extend the pseudorapidity $\eta$ coverage provided by the barrel and
endcap detectors. Muons are measured in gas-ionization detectors embedded in the
steel flux-return yoke outside the solenoid. A more detailed description of the
CMS detector, together with a definition of the coordinate system used and the
relevant kinematic variables, can be found in Ref.~\cite{CMS:2008xjf}.

The silicon tracker used in 2016 measured charged particles within the range
$\abs{\eta} < 2.5$. For nonisolated particles with the transverse momentum of $1 < \pt < 10\GeV$ and
$\abs{\eta} < 1.4$, the track resolutions were typically 1.5\% in \pt and 25--90
(45--150)\mum in the transverse (longitudinal) impact
parameter~\cite{CMS:2014pgm}. At the start of 2017, a new pixel detector was
installed~\cite{Phase1Pixel}; the upgraded tracker measures particles up to
$\abs{\eta} = 3.0$ with typical resolutions of 1.5\% in \pt and 20--75\mum in
the transverse impact parameter~\cite{DP-2020-049} for nonisolated particles with
$1 < \pt < 10\GeV$.

Muons are measured in the range $\abs{\eta} < 2.4$, with
detection planes made using three technologies: drift tubes, cathode strip
chambers, and resistive-plate chambers. Matching muons to tracks measured in the
silicon tracker results in a relative \pt resolution, for muons
with \pt up to 100\GeV, of 1\% in the barrel and 3\% in the
endcaps~\cite{CMS:2018rym}.

Events of interest are selected using a two-tiered trigger system. The first level, 
composed of custom hardware processors, uses information from the calorimeters 
and muon detectors to select events at a rate of around 100\unit{kHz} within a fixed 
latency of about 4\mus~\cite{CMS:2020cmk}. The second level, known as the 
high-level trigger, consists of a farm of processors running a version of the 
full event reconstruction software optimized for fast processing, and reduces the 
event rate to around 1\unit{kHz} before data storage~\cite{CMS:2016ngn}.

\section{Data and Monte Carlo simulation}
\label{sec:data}

We split the data collected with the CMS detector in 2016--2018 into four
distinct periods: 2016a, 2016b, 2017, and 2018. The integrated luminosities of the corresponding samples
are 20.0, 16.6, 42.0, and 61.3\fbinv~\cite{CMS:2021xjt,LUM-17-004,LUM-18-001}.
Data from the 2016a period were affected by strip tracker dynamic hit inefficiency. The problem
was resolved in August 2016 and the rest of the 2016 data-taking period
is referred to as 2016b. Before the 2017 data-taking period, the pixel detector was upgraded, improving the acceptance, redundancy, and resolution. During data taking in 2017,
the pixel detector had synchronization issues at the beginning of the
run, followed by DC-DC converter failures~\cite{Phase1Pixel} leading to about 11\% of the
detector being unresponsive. Most of these issues were resolved before data taking began in 2018.

We use multiple samples of MC simulated events to evaluate the signal efficiency, 
the detector response, and the background yields. The simulated event samples are generated with
\PYTHIA\ 8.212~\cite{Sjostrand:2014zea} using the CP5 underlying event tune~\cite{CMS:2019csb} and
propagated through the CMS detector model using the
\GEANTfour~\cite{Agostinelli:2002hh} package. The decays of \PB hadrons are
simulated using the \EVTGEN 1.3.0~\cite{Lange:2001uf} program and final-state photon
radiation is described using \PHOTOS 3.56~\cite{Davidson:2010ew}.

Multiple interactions within the same or nearby bunch crossings (pileup) are simulated for all samples by
overlapping the hard-scattering event with several minimum bias events, with the multiplicity similar to the
one observed in data (averaging to 23 for 2016 and 32 for 2017--2018).

\section{Event reconstruction and selection}
\label{sec:selection}

The events used in this analysis were collected with a set of dimuon triggers
designed to select \bmm, \bjpsik, and \bjpsiphi events. To achieve an acceptable
trigger rate, the first-level trigger required two
high-quality~\cite{CMS:2020cmk} oppositely charged muons restricted
to $\abs{\eta}<1.5$. At the high-level trigger, a high-quality dimuon secondary
vertex (SV)~\cite{CMS:2016ngn} was required and the events were restricted to
mass ranges of 4.5--6.0\GeV and 2.9--3.3\GeV for the \PB and \PJGy mesons,
respectively. The \PJGy triggers additionally required the SV
to be displaced from the beam spot (defined as the average
interaction point in the plane transverse to the beams) and the displacement vector to be aligned with the
dimuon momentum.

The event candidate selection starts with the reconstruction of dimuon candidates,
which are used to build different \PB mesons for the signal and normalization channels.
The selection is kept as similar as possible for all the channels
to benefit from the cancellation of systematic effects in the ratio
of the efficiencies used for the branching fraction normalization. Both muons are
required to have a high-quality inner track~\cite{CMS:2014pgm} with
$\pt>4\GeV$ and $\abs{\eta}<1.4$, and pass the tight muon identification
requirements~\cite{CMS:2018rym, CMS:2019bbr}, which suppress misidentified
muons from pion and kaon decays, and from other sources.
Extra single and double kaons with $\pt>1\GeV$ and $\abs{\eta}<2.5$ are required for the \bjpsik and \bjpsiphi normalization channels, respectively.

We obtain \PB candidate properties by employing the kinematic fitter described
in Ref.~\cite{Prokofiev:2005zz}. We apply different kinematic constraints depending
on the final state. For the \bsmm and \bdmm decays, we use the vertex constraint, while
for the \bjpsik and \bjpsiphi decays we also add a mass constraint for the \PJGy candidate.

From the \PB candidate's decay vertex and its momentum we build a refitted
trajectory representing the \PB candidate. Then, for each reconstructed primary vertex (PV) in
the event, the trajectory is extrapolated to the point closest to that
vertex. The absolute distance between the closest point and the PV in 3D space is defined as the
impact parameter. The PV with the smallest impact parameter is selected as the best PV for 
the \PB candidate.

Table~\ref{tab:candidate_selection} shows a summary of the \PB candidate
selection requirements for the signal extraction and normalization UML fits.
The 3D SV displacement significance is defined as the 3D distance between the PV and the
dimuon SV, divided by its corresponding uncertainty. The 2D \MM pointing
angle $\alpha$ is defined as the angle between the dimuon momentum and the line
connecting the PV and SV, and is calculated in the transverse plane with respect
to the beam direction. The pointing angle requirement is introduced to match
the $\cos\alpha>0.9$ requirement used in the \PJGy triggers.

\begin{table*}[h]
  \centering
  \topcaption{Selection requirements for the three decay channels
    used in the signal yield and normalization fits. Addition
    selection requirements are applied for the \bjpsik control sample
    used in systematic studies.}
      \begin{tabular}{p{7cm}ccc}
        Selection                                    & \bmm          & \bjpsik       & \bjpsiphi\\
        \hline
        \PB candidate mass [\GeVns{}]                & [4.90, 5.90]  & [4.90, 5.90]  & [4.90, 5.90]\\
        Blinding window [\GeVns{}]                   & [5.15, 5.50]  &               &  \\
        \ptmu  [\GeVns{}]                            & $>$4          & $>$4          & $>$4\\
        $\abs{\etamu}$                               & $<$1.4        & $<$1.4        & $<$1.4\\
        3D SV displacement significance              & $>$6          & $>$4          & $>$4\\
        \MM \pt  [\GeVns{}]                          & $>$5          & $>$7          & $>$7\\
        \MM SV probability                           & $>$0.025      & $>$0.1        & $>$0.1\\
        \MM invariant mass [\GeVns{}]                &               & [2.9, 3.3]    & [2.9, 3.3]\\
        Kaon \pt [\GeVns{}]                          &               & $>$1          & $>$1\\
        \PJGy mass-constrained fit probability       &               & $>$0.025      & $>$0.025\\
        2D \MM pointing angle [rad]                  &               & $<$0.4        & $<$0.4\\
        $\PGfP{1020}$ candidate mass [\GeVns{}]      &               &               & [1.01, 1.03]\\
        \mvab                                        & $>$0.9        &               & \\
      \end{tabular}
    \label{tab:candidate_selection}
\end{table*}

\section{Multivariate analysis}
\label{sec:mva}
Most of the observables used to distinguish signal from background have rather
weak discriminating power. Therefore, we employ an MVA to
combine them into a single, more powerful discriminator \mvab. Compared to
the previous analysis~\cite{CMS:2019bbr}, we have relaxed the preselection requirements, developed
new discriminating observables, added significantly more data for the model
training, and used a more advanced machine learning algorithm. This allows us to
significantly improve the analysis sensitivity, achieving the same level as in the
previous measurement with just $\sim$60\% of the previous data.

Inputs to the MVA are split into three major classes. The
first class includes pointing angles, which are defined as the angles between the
\PB candidate momentum and the line connecting the PV and SV, either in 2D or 3D. 
We use both definitions since the 2D version benefits from a smaller uncertainty in the vertex position and the 3D version
provides additional matching information along the beam line. These observables
are effective at rejecting all types of backgrounds, except for the ones originating from the
two-body decays.

The second class of observables is related to the SV. The dimuon
candidates from the combinatorial background tend to be associated with a low-quality
SV fit. Therefore, the SV probability, calculated using the $\chi^{2}$ and the number of degrees of freedom of the fit,
is one of the most powerful
discriminating variables. Furthermore, we achieve additional background suppression by
rejecting events that contain a better-quality SV formed by one of the muons and any track in the event.
Finally, most of the misreconstructed SVs tend to be close to the PV and therefore can be
rejected using the SV displacement information.

The last class of observables is designed to detect nearby decay products
present in semileptonic decays of \PQb and \PQc hadrons. 
We compute the number of tracks compatible with the \MM SV. 
In addition, isolation variables are calculated for the \PB candidate as well as both muon candidates.
The isolation observables have been optimized in the previous
studies~\cite{CMS:2019bbr} to maximize the separation power between the \bsmm
signal and the background, while maintaining a reasonable agreement between the
MC simulation and the data for the \bjpsik normalization channel. The isolation
is defined as
\begin{linenomath}
\begin{align*}
 I = \frac{\pt}{\pt + \sum_\text{trk}\pt(\text{trk})},
\end{align*}
\end{linenomath}
where \pt is the momentum of either the \PB or muon candidate and the sum includes
all charged-particle tracks with $\pt>0.9\GeV$ in a cone of radius
$\Delta R = \sqrt{\smash[b]{(\Delta\eta)^2 + (\Delta\varphi)^2}}$
around the candidate momentum direction, with $\varphi$ representing the
azimuthal angle in radians. Only the charged particles that do not belong to
the candidate and are associated with the same PV are considered. The $\Delta R$
is required to be smaller than 0.7 for the \PB candidate isolation and 0.5 for
the muon isolation.

For the MVA training, we employ the \textsc{XGBoost}
library~\cite{Chen:2016:XST:2939672.2939785}, which implements an advanced
gradient boosting algorithm. The training is performed on a mixture of simulated \bsmm signal events and
background events in data selected using the sidebands of the dimuon mass
distribution consisting of two regions: 5.5--5.9\GeV populated by the
combinatorial background and 4.9--5.1\GeV representing a combination of the
partially reconstructed and combinatorial backgrounds. The events are split into
training and testing categories in a 2:1 proportion. To reuse all available
events, we train three classifiers by assigning events to one of the categories
based on their event number modulo 3. This allows us to classify all events in
data, making sure that no event was evaluated by a classifier that was trained
on the event itself.

The \bjpsik selection requirements in Table~\ref{tab:candidate_selection} define
the \textit{normalization} sample used to extract the final branching fractions. We
also define a \bjpsik \textit{control} sample. This sample starts with an
additional selection requirement of kaon $\pt < 1.5\GeV$, which effectively requires
the kaon to carry only a small fraction of the parent \PB meson momentum,
providing a better match with the \bmm kinematic distributions.

Even with the kaon $\pt < 1.5\GeV$ requirement, there are still differences in the
kinematic distributions for the two
decays, which may have an impact on the analysis. The most important one is the
difference in the invariant mass of the \Bs and \PJGy mesons, which has a significant impact
on the opening angle between the two muons. The \PJGymm decay tends to have a larger uncertainty
in the \MM vertex position along the dimuon momentum. Therefore the
SV significance for the \bjpsik events needs to be scaled by a factor of $\sim$1.6 to
match the \bsmm distribution.

We use the \bjpsik control sample to measure the MVA performance in data.
To achieve the best matching between the \mvab distributions for the \bsmm and \bjpsik
decays, we need to select appropriate input observables for the MVA in the \bjpsik
control sample. For the pointing angle and the impact parameter we use the $\MM \PKp$ final-state
observables, since otherwise we would have an incorrect \PB candidate momentum vector. We also
ignore the kaon track in all the isolation calculations and extra track counting. For
the remaining inputs, we rely on the \MM observables only.

\begin{figure*}[!htb]
\centering
    \includegraphics[width=0.32\textwidth]{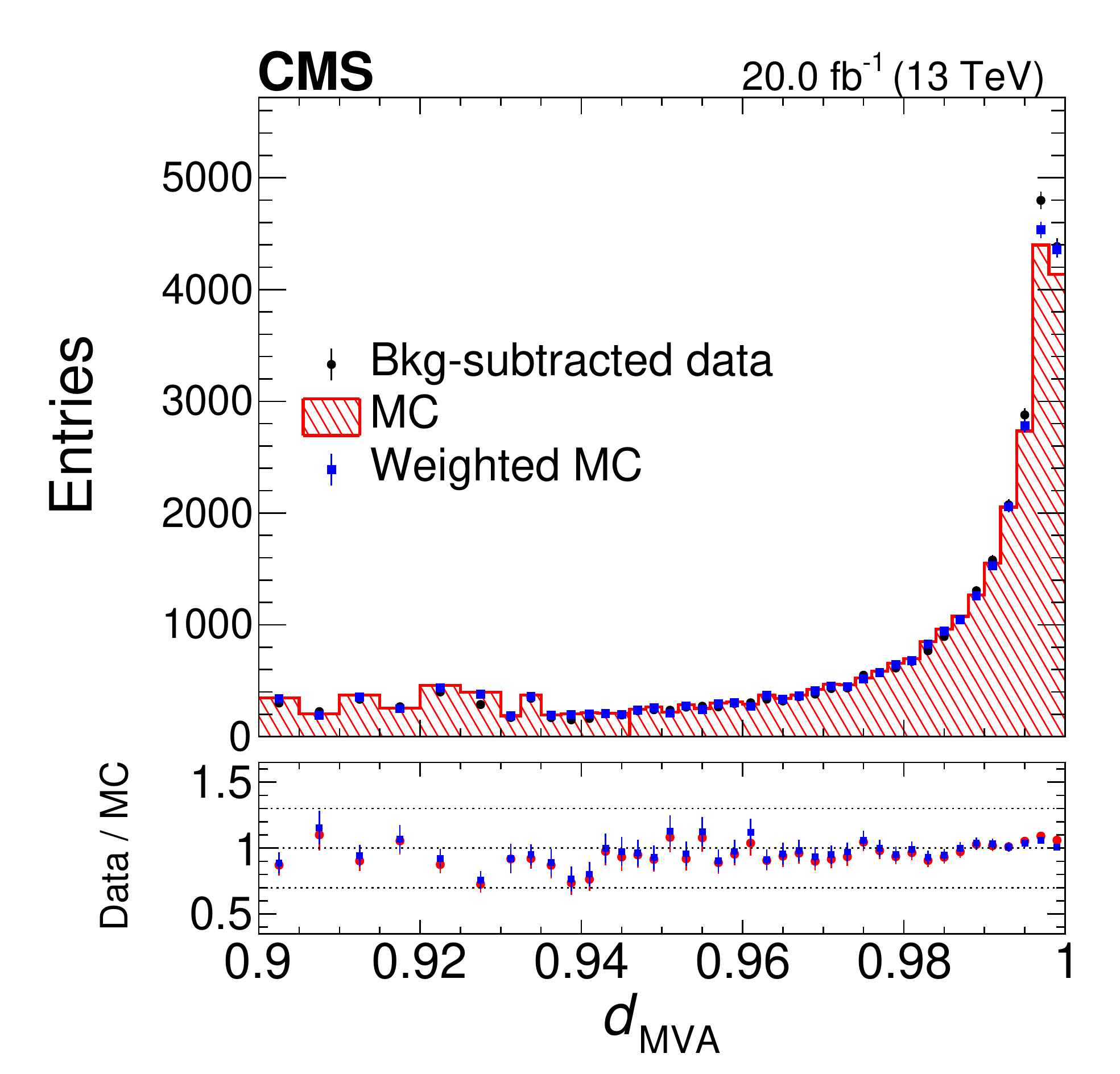}
    \includegraphics[width=0.32\textwidth]{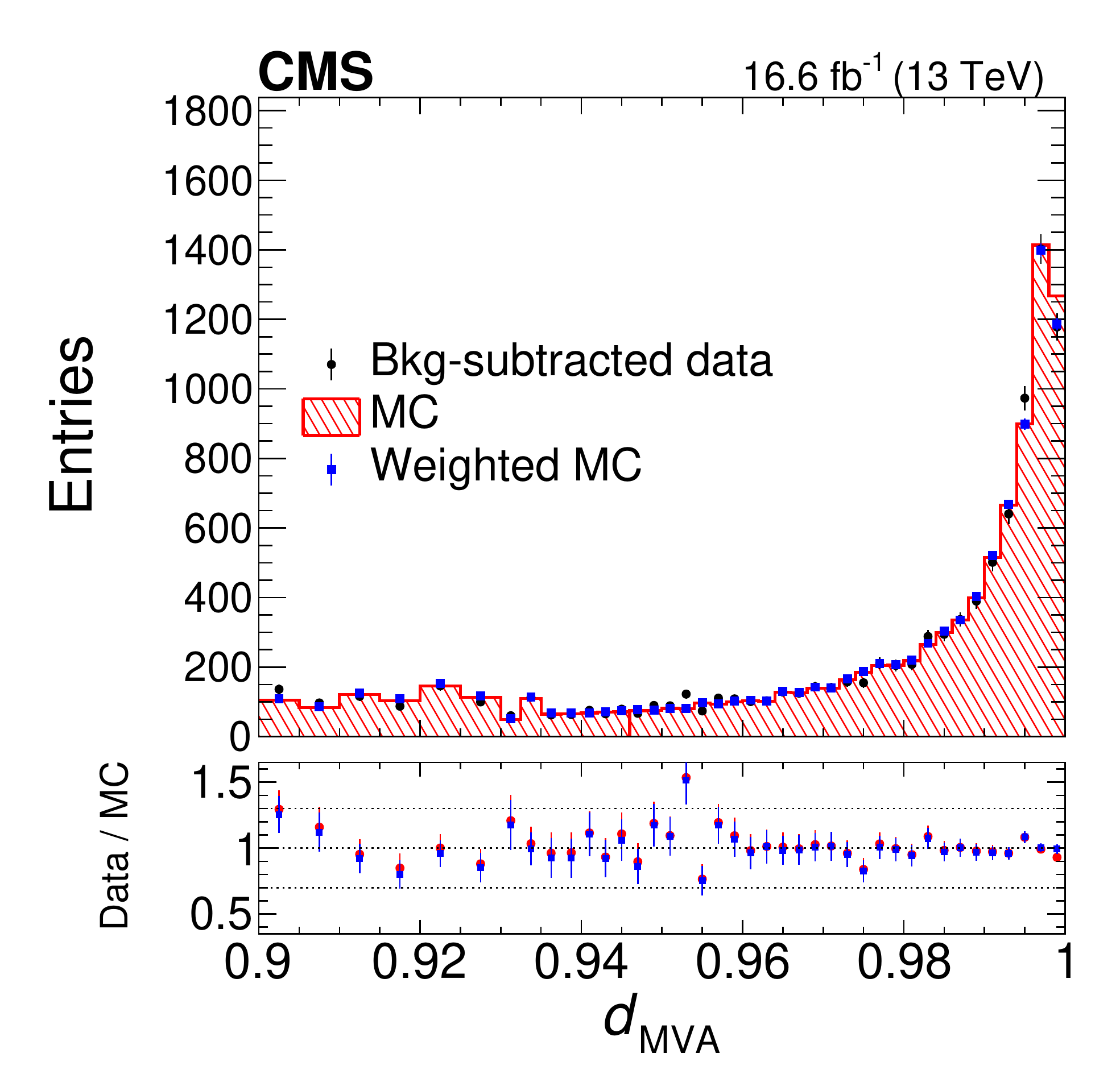}
    \includegraphics[width=0.32\textwidth]{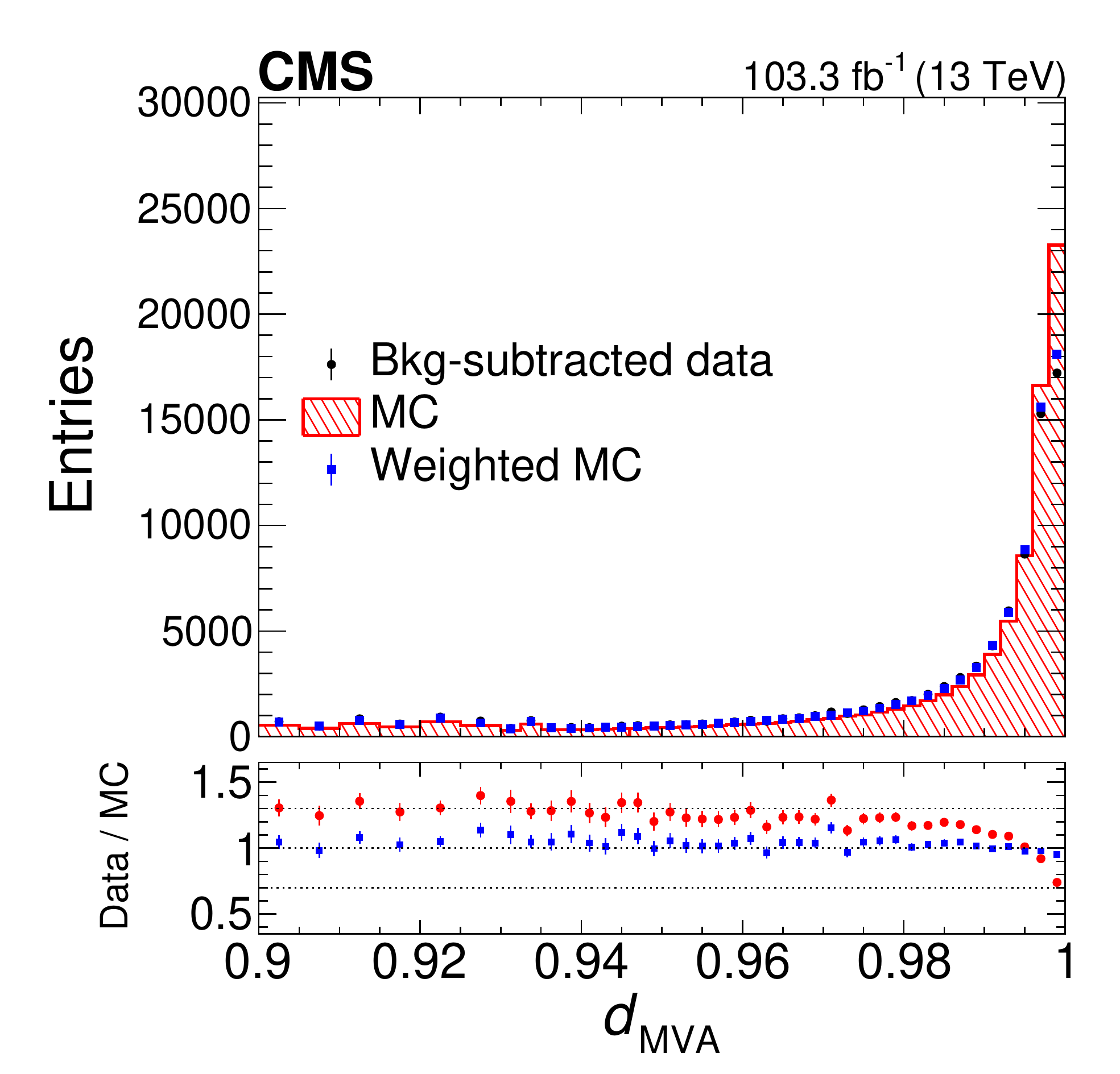}\\
  \caption{Distributions of the \mvab output for the 2016a (left), 2016b
    (center), and 2017--2018 (right) data and the corresponding MC samples. The
    blue squares represent the weighted simulated distributions using the
    \textsc{XGBoost} reweighting method. In the lower panel, the blue squares
    and red points are the ratio of the data to weighted and not weighted
    simulated distribution respectively.  The MC distributions are normalized to
    the total number of events in data.}
  \label{fig:jpsik_data_mc_summary}
\end{figure*}

Figure~\ref{fig:jpsik_data_mc_summary} shows a comparison of the \mvab
distributions for the \bjpsik simulation and data. The data plots have 
backgrounds subtracted using the \textit{sPlot} technique~\cite{Pivk:2004ty} applied to the
UML fits of the \bjpsik invariant mass distributions. We observe good agreement between 
the MC simulation and data for the 2016a and 2016b samples. The agreement is worse for 
the 2017 and 2018 samples.

We derive corrections to the efficiency of the \mvab selection requirements, defined in the caption of
Table~\ref{tab:mva_corrections}, in two different ways. The first (``Ratio") method derives the corrections using
the \bjpsik data and MC samples and applies them to the \bsmm and \bdmm efficiencies. The
second (``\textsc{XGBoost}'') method is based on the idea of reweighting the MC
simulation samples to match the data. We were not able to find a single
variable that would allow us to compensate for the discrepancy. Therefore
we developed an approach using the \textsc{XGBoost} algorithm to train
a classifier on the difference between the simulation and data in \bjpsik events
and use it to reweight the simulated \bmm events. We trained the \textsc{XGBoost}
classifier using the same inputs that we use for the MVA.
The backgrounds are subtracted from the data via the \textit{sPlot} technique, as described
above. The corrections from the two methods are summarized in
Table~\ref{tab:mva_corrections}. In general, the two methods give
results compatible within 1--2$\sigma$. We use the results from the \textsc{XGBoost}
method as a default, and take the difference between the results of the two methods as a systematic
uncertainty.

\begin{table*}[h]
\centering
    \topcaption{Efficiency corrections for the \bsmm decays derived using two different
    methods: the efficiency ratio between data and simulation, and the \textsc{XGBoost} reweighting 
    in \bjpsik events. The quoted uncertainties are statistical only.}
      \label{tab:mva_corrections}
        \cmsTable{
        \begin{tabular}{lcccccc}
          \multirow{2}{*}{Method} & \multicolumn{3}{c}{$\mvab>0.9$ selection} & \multicolumn{3}{c}{$\mvab>0.99$ selection} \\
                            & 2016            & 2017            & 2018            & 2016 & 2017 &	2018 \\
          \hline
          Ratio       		    & $1.011\pm0.013$ & $0.939\pm0.007$ & $0.903\pm0.008$ & $1.058\pm0.019$ & $0.891\pm0.008$ & $0.885\pm0.010$ \\
          \textsc{XGBoost}         & $0.991\pm0.008$ & $0.949\pm0.003$ & $0.917\pm0.002$ & $1.008\pm0.011$ & $0.905\pm0.004$ & $0.908\pm0.002$ \\
        \end{tabular}
        }
\end{table*}

\section{Data analysis}
\label{sec:fitting}
In order to extract results we perform a set of UML fits:
the \bjpsik and \bjpsiphi yield fits, the simultaneous \bsmm and \bdmm branching
fraction fit, and the \bsmm effective lifetime fit.

For the branching fraction measurement, we perform a 2D fit of the dimuon invariant mass and
the relative mass resolution distributions within multiple event categories. The
events are categorized using the following three independent criteria:
\begin{itemize}
\item[--] data-taking period: 2016a, 2016b, 2017, or 2018,
\item[--] signal purity based on the \mvab value: 0.90--0.99 or 0.99--1.00,
\item[--] $\abs{\eta}$ of the most forward muon: 0.0--0.7 or 0.7--1.4,
\end{itemize}
leading to 16 distinct categories. The separation by $\abs{\eta}$ of the most forward
muon is motivated by different signal purities and mass resolutions in the two
regions. The parameters of interest are the branching fractions of the \bsmm and \bdmm
decays, which are derived from the corresponding yields.

The likelihood consists of five components: the \Bs signal, the \Bd signal, the
partially reconstructed semileptonic background, the peaking \bhh background,
where \Ph represents a hadron, and the combinatorial background. 
The statistical uncertainties from the MC simulation and the systematic 
uncertainties are propagated to the final results by introducing nuisance 
parameters, which are profiled during the fit.

The signal components are modeled by Crystal Ball functions~\cite{CB} and
include the per-event mass resolution in the parameterization. The mass
resolution and scale are calibrated using the $\PJGy \to \MM$ and $\PGU(1S) \to
\MM$ data samples. All the parameters of the signal model are fixed in the fit, except for the
width of the Crystal Ball function, which is a conditional parameter proportional to the dimuon mass resolution.

The semileptonic background is dominated by the three-body $\PB\to\Ph^-\PGmp\PGn$ and
$\PB\to\Ph\MM$ processes. The contribution of this component is determined from the simulation of $\PB^0\to\PGpm\PGmp\PGn$, $\Bs\to \PKm\PGmp\PGn$, $\PB^+\to\PGpp\MM$, and 
$\PB^0\to \PGpz\MM$ decays. Contributions from $\PgL_{\PQb}\to\Pp\PGmm\PGn$ 
and $\PB_{\PQc}^+\to\PJGy\PGmp$ backgrounds are found to be negligible
compared with the uncertainty in the dominant background normalization.
The shape of the semileptonic background is a Gaussian function with the mass 
and width parameters allowed to vary in the fit to the data.

The peaking \bhh background is represented by a sum of Gaussian and Crystal Ball 
functions with a common central mean value. The parameters used in the fit to the 
data are fixed from the results of a fit to the distribution obtained from MC simulation of 
$\Bd\to \PKp\PKm$, $\PKp\PGpm$, $\PGpp\PGpm$, and $\Bs\to \PKp\PKm$, $\PKm\PGpp$, $\PGpp\PGpm$ decays,
with branching fractions from Ref.~\cite{Workman:2022ynf}.

The yields of the semileptonic and peaking background components are calculated
using the normalization channel and the corresponding MC simulation with the
corrected efficiencies. Systematic uncertainties of 25\% and 50\% are assigned 
to the semileptonic and peaking backgrounds, respectively, to account for uncertainties in 
the rate of misidentifying charged hadrons as muons.
The nuisance parameters in the yield calculation, such
as the ratio of the efficiencies and the normalization of the \bjpsik process, are constrained
using Gaussian or log-normal priors, according to the corresponding systematic uncertainties.

The combinatorial background is modeled by a linear function (constrained to be
positive in the entire fit range). The yields and the slopes of the
combinatorial background are free parameters in the UML.

We estimate the expected performance of the branching fraction measurement via
an ensemble of pseudo-experiments generated using the SM values for the
branching fractions and the lifetime. The relative uncertainties, 
which include the systematic uncertainties described in Section~\ref{sec:systematics},  
are expected
to be ${}^{{+}11.1}_{{-}10.5}\,\%$ in \cbf(\bsmm) and ${}^{{+}67}_{{-}62}\,\%$ in $\cbf(\bdmm)$.

To extract the effective lifetime of the \Bs meson in the \bsmm decay,
we perform a 3D UML fit to the dimuon invariant mass, decay time, and decay time
uncertainty, dividing the data by data-taking period, \mvab value,
and rapidity of the most forward muon, as we do for the branching fraction fit.
The signal acceptance as a function of the decay time is extracted from simulation and
corrected with the \bjpsik decays in data.
To minimize the differences between the two channels, we use the \bjpsik
control sample defined in Section~\ref{sec:mva}, along with its MVA.
The combinatorial background decay time distribution was
obtained from a mass sideband in data. The decay time uncertainty is
calculated for each event and is used as an observable in the fit.
The decay time uncertainty models are obtained from the simulation samples and sideband data.
Using pseudo-experiments generated with a complete \bsmm model, the expected
total uncertainty in the lifetime is found to be ${}^{{+}0.18}_{{-}0.16}\unit{ps}$.

\section{Systematic uncertainties}
\label{sec:systematics}
\subsection{Branching fraction measurement}
The branching fraction measurements have multiple sources of experimental and
theoretical systematic uncertainties. The experimental uncertainties are
dominated by the uncertainty in the \bmm signal efficiency corrections due to
mismodeling of \mvab in MC simulation, the kaon reconstruction and selection
efficiency for the \bjpsik and \bjpsiphi normalization measurements, and the
trigger efficiency difference between the signal and normalization channels.
The uncertainties in the branching fractions of the \bjpsik and \bjpsiphi
decays, as well as in \fsfu, are considered to be external uncertainties, which
are factorized out in the final results.

The signal efficiency corrections for mismodeling of the \mvab distribution are
estimated with two different methods described in Section~\ref{sec:mva}. 
The two methods give results compatible with each other. Based on the
difference between the two methods, we assign a 2 (3)\% systematic uncertainty 
in the corrections for the \bsmm and \bdmm signal efficiencies for the $\mvab>0.90$ (0.99) selection (``\mvab correction'').

The hadron tracking efficiency uncertainty is obtained by comparing the ratio of the measured
branching fractions of the two-body $\PDz \to \PKm\PGpp$ to the four-body
$\PDz \to \PKm\PGpp\PGpp\PGpm$ decays to the world average value~\cite{Workman:2022ynf}.
This gives a ``Tracking efficiency'' uncertainty of 2.3\%~\cite{DP-2022-012} for each kaon.

As a result of the different kinematics and triggers for the signal and normalization channels, 
the trigger efficiency effects do not fully cancel and are corrected using MC simulation. 
The simulation of the trigger efficiency is checked by comparing with the efficiency measured 
using data obtained from other triggers. The observed differences between simulation and data 
are used to establish a ``Trigger efficiency'' systematic uncertainty of 3.7 (2.4)\% for 2016 (2017--2018).

There are a few more systematic uncertainties in the measurement. The fit bias is extracted from the difference 
between the branching fraction obtained from the pseudo-experiments and the input SM value,
combined with the variation caused by using different background models in the fit (``Fit bias'').
The shape uncertainty in the normalization channel is derived by using different 
signal templates in the yield fits (``\bjpsik shape uncertainty''). The pileup uncertainty is extracted from the difference 
in the efficiency performance derived using the pileup distribution in data and in MC simulation (``Pileup''). 
The normalization channels use a tighter SV probability requirement than the signal channel 
because of the different triggers. The corresponding uncertainty is evaluated by
comparing the efficiency difference between the tighter and the signal channel
SV probability requirement ($>$0.100 with respect to $>$0.025) in the data 
and MC simulation in \bjpsik events (``Vertex quality requirement''), where
the data is from a sample that is triggered without the SV probability requirement.

Table~\ref{tab:sys_bsmm} summarizes the systematic uncertainties for
the branching fraction measurements using \bjpsik events for
normalization, including the \bjpsik branching fraction uncertainty~\cite{Workman:2022ynf}.
For the \bsmm branching fraction measurement with the \bjpsiphi
normalization, the systematic uncertainty in the tracking efficiency is doubled to 4.6\% due to
the presence of two kaons in the final state, and the shape uncertainty is found to
be 1.5\%. At the same time, this measurement could be free from explicitly taking into account the
\PB production fraction systematic uncertainty, as discussed in Section~\ref{sec:results}.

The lifetime of the \Bs meson has a significant impact on the signal efficiency
for the \bsmm decays. The branching fraction measurements are reported assuming
the SM value for the lifetime (1.61\unit{ps}). Since the lifetime affects the branching 
fraction measurements, we provide a correction for alternative 
lifetime hypotheses. The scale factor for the branching fraction is $1.577 - 0.358\tau$, 
where $\tau$ is the \Bs meson lifetime in ps.

\begin{table}[thp]
\centering
 \topcaption{Summary of the systematic uncertainties for the \bsmm and \bdmm
  branching fraction measurements.}
   \tableCMS{
    \begin{tabular}{lcc}
      Effect & \bsmm & \bdmm\\
      \hline
      \fsfu ratio of the \PB meson production fractions   & 3.5\% & \NA  \\
      \mvab correction                  &  \multicolumn{2}{c}{2--3\%} \\
      Tracking efficiency (per kaon)    &  \multicolumn{2}{c}{2.3\%} \\
      Trigger efficiency                &  \multicolumn{2}{c}{2.4--3.7\%} \\
      Fit bias                          &  2.2\%   & 4.5\% \\
      Pileup                            &  \multicolumn{2}{c}{1\% } \\
      Vertex quality requirement        &  \multicolumn{2}{c}{1\%} \\
      \bjpsik shape uncertainty         &  \multicolumn{2}{c}{1\%} \\
      \bjpsik branching fraction        &  \multicolumn{2}{c}{1.9\%} \\
    \end{tabular}}
    \label{tab:sys_bsmm}
\end{table}

\subsection{Lifetime measurement}
The dominant source of systematic uncertainty in the lifetime measurement is
associated with mismodeling of the correlation between the \mvab and the decay
time. This correlation stems from the most discriminating MVA variables, the pointing angle and its uncertainty, 
both of which are strongly
correlated with the decay time. The correlation enters via the decay distance:
the larger the decay distance is, the better one knows the direction from the PV
to SV. As the decay distance gets shorter, the uncertainty in the pointing
angle increases, making such events harder to distinguish from the
background. Mismodeling of these correlations in simulation can have a
significant impact on the decay time distribution.

The decay time is also correlated with many selection requirements. Most of them are
well simulated. We measure the lifetime bias in \bjpsik events using a relaxed
selection requirement, $\mvab>0.90$, and compare it to the prediction from
simulation. We find the bias for the lifetime measurement to be
0.04--0.05\unit{ps}, depending on the data-taking period (``Lifetime fit bias").

For the final selection, we derive a correction as a ratio of the decay time
distributions for the $\mvab>0.99$ and $\mvab>0.90$ requirements using \bjpsik events in data. Then we 
apply this correction to the \bsmm decay time distribution extracted from 
simulation using $\mvab>0.90$ as the selection requirement. Repeating the procedure using
simulated events, we find that the method may introduce a bias of up to 0.10\unit{ps}
in 2016 data. The bias is found to be much smaller in 2017 and 2018 data. These
effects (``Decay time distribution mismodeling'') are taken into account in the
lifetime fit by introducing independent
nuisance parameters in the fit model.

Two additional minor systematic uncertainties are also included. The uncertainty from the imperfect
parameterization of the efficiency dependence on the decay time is 
derived using different analytical functions in the lifetime 
fit to \bjpsik events (``Efficiency modeling''). We also measure the lifetime in the MC samples generated with 
different lifetimes from the pseudo-experiments while sharing the same efficiency 
function. The difference between the measured lifetime and the input lifetime of 
the MC samples is assigned as a systematic uncertainty (``lifetime dependence").

Table~\ref{tab:sourcesys} summarizes the systematic uncertainties in the
lifetime measurement. The uncertainties of 2017 and 2018 data-taking periods
are treated as correlated
and other two are treated as uncorrelated.

\begin{table}[thp]
\centering
  \topcaption{Summary of the systematic uncertainties in the \bsmm
    effective lifetime measurement (in ps) in four data-taking periods.}
    \tableCMS{
      \begin{tabular}{ lcccc}  
        Effect & 2016a  & 2016b & 2017 & 2018\\
        \hline
        Lifetime fit bias     & 0.04 & 0.04 & 0.05 &0.04\\
        Decay time distribution mismodeling  & 0.10 & 0.06 & 0.02 &0.02\\
        Efficiency modeling & \multicolumn{4}{c}{0.01}\\
        Lifetime dependence & \multicolumn{4}{c}{0.01}\\[\cmsTabSkip]
        Total & 0.11 & 0.07 & 0.05 & 0.04\\
      \end{tabular}}
  \label{tab:sourcesys}
\end{table}

\section{Results}
\label{sec:results}
Using the result of the \bjpsik normalization fit with
Eqs.~\eqref{eq:bf_bs_jpsik} and~\eqref{eq:bf_bd}, we find the branching
fractions to be:
\begin{linenomath}
\ifthenelse{\boolean{cms@external}}
{ 
\begin{multline*}
\cbf(\bsmm) = \\ =\left[ 3.83^{+0.38}_{-0.36}~\stat\,^{+0.19}_{-0.16}\syst\,^{+0.14}_{-0.13}\,\text{(\fsfu)} \right] \times 10^{-9}, 
\end{multline*}
\begin{equation*}
\cbf(\bdmm) = \left[ 0.37^{+0.75}_{-0.67}~\stat\,^{+0.08}_{-0.09}\syst \right] \times 10^{-10}.
\end{equation*}
} 
{ 
\begin{align*}
  \cbf(\bsmm) 
  &= \left[ 3.83^{+0.38}_{-0.36}~\stat\,^{+0.19}_{-0.16}\syst\,^{+0.14}_{-0.13}\,\text{(\fsfu)} \right] \times 10^{-9}, \\
  \cbf(\bdmm) 
  &= \left[ 0.37^{+0.75}_{-0.67}~\stat\,^{+0.08}_{-0.09}\syst \right] \times 10^{-10}.
\end{align*}
} 
\end{linenomath}

\begin{figure}[tp]
\centering
    \includegraphics[width=0.48\textwidth]{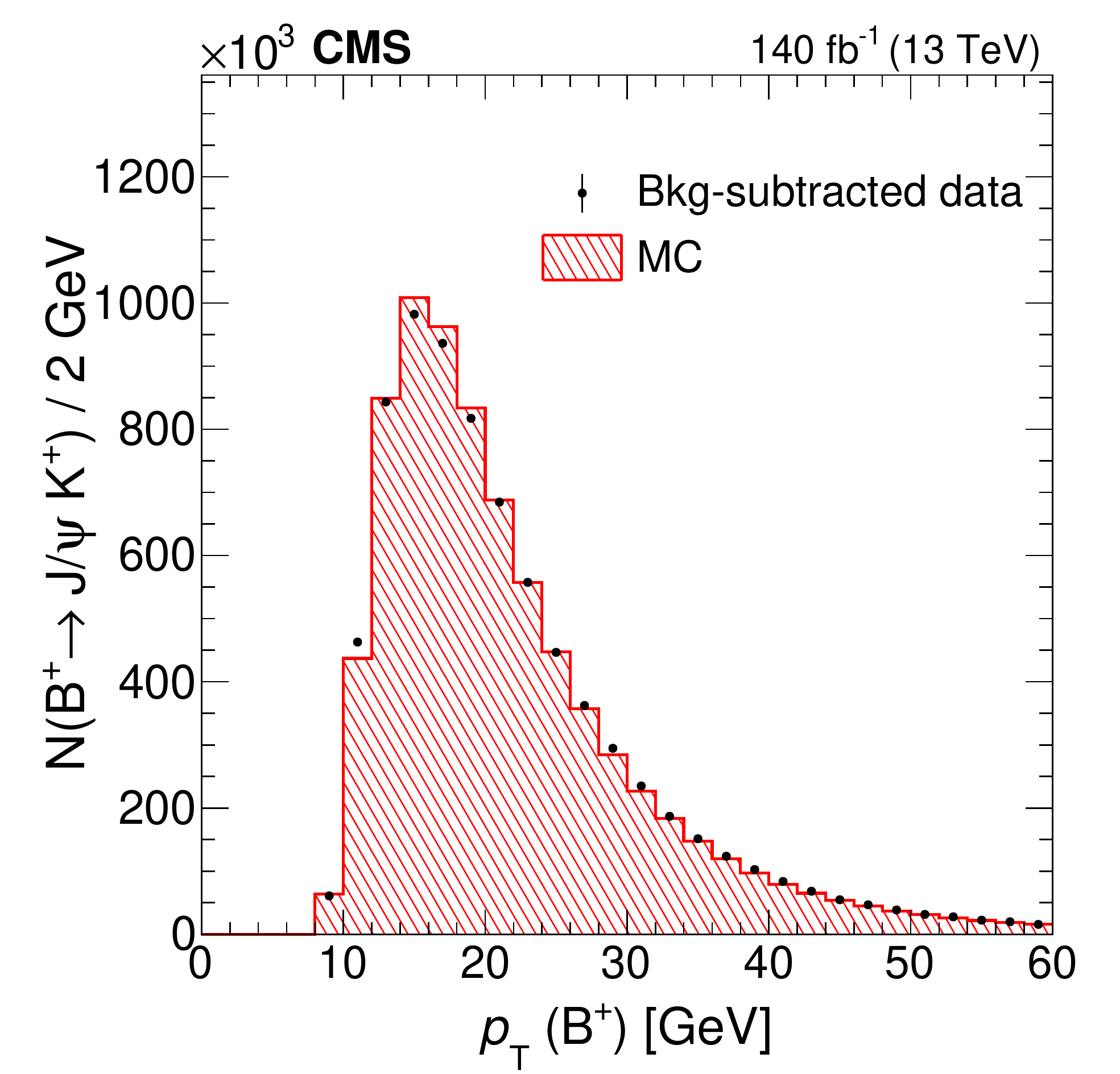}
    \includegraphics[width=0.48\textwidth]{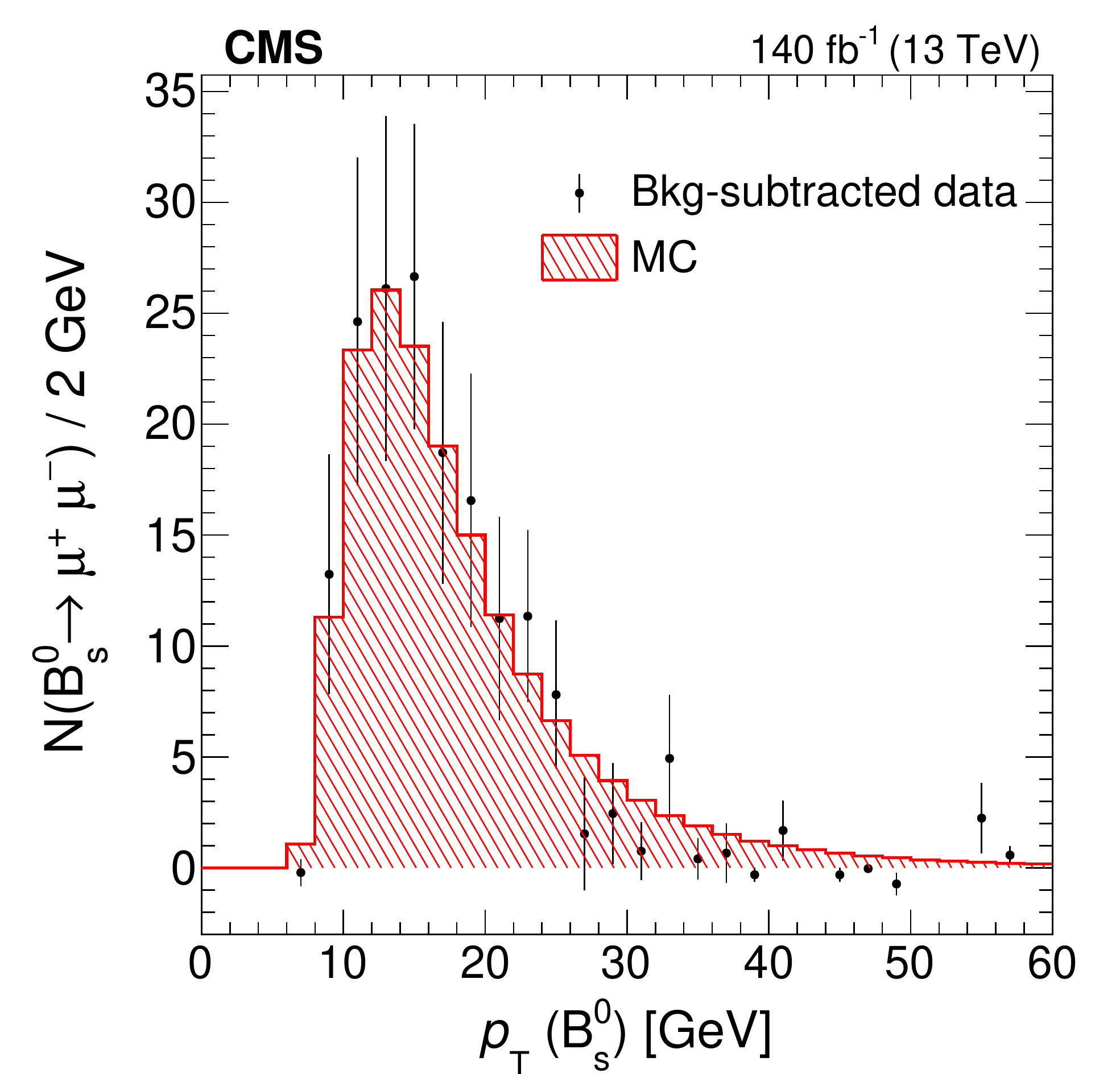}
    \caption{The distribution of the \PB meson \pt after the \textit{sPlot}
      background subtraction in data (points with error bars) and simulation
      (hatched histogram) for \bjpsik (\cmsLeft) and \bsmm (\cmsRight) events. The
      MC distributions are normalized to the total number of events in data.}
    \label{fig:bpt_splot}
\end{figure}

The correlation between the extracted $\cbf(\bsmm)$ and $\cbf(\bdmm)$ branching fractions is $-0.120$.
These results are based on the following external inputs:
\begin{itemize}
\item $\cbf(\bjpsik) = (1.020\pm0.019) \times 10^{-3}$,
\item $\cbf(\PJGymm) = (5.961\pm0.033) \times 10^{-2}$, and
\item $\fsfu =0.231\pm0.008$.
\end{itemize}
The branching fractions are taken from
Ref.~\cite{Workman:2022ynf}. The \fsfu ratio is derived from the
\pt-dependent measurement of the \fsfu ratio by LHCb~\cite{LHCb:2021qbv}. We are
using the \pt\ distribution observed in this measurement, shown in Fig.~\ref{fig:bpt_splot},
to compute an effective \fsfu ratio for the corresponding phase space. 
Measurements of the \pt and $\eta$ dependence of the ratio by the CMS
Collaboration~\cite{CMS:2022wkk} are found to be consistent with the LHCb results.

\begin{figure}[tp]
\centering
    \includegraphics[width=0.48\textwidth]{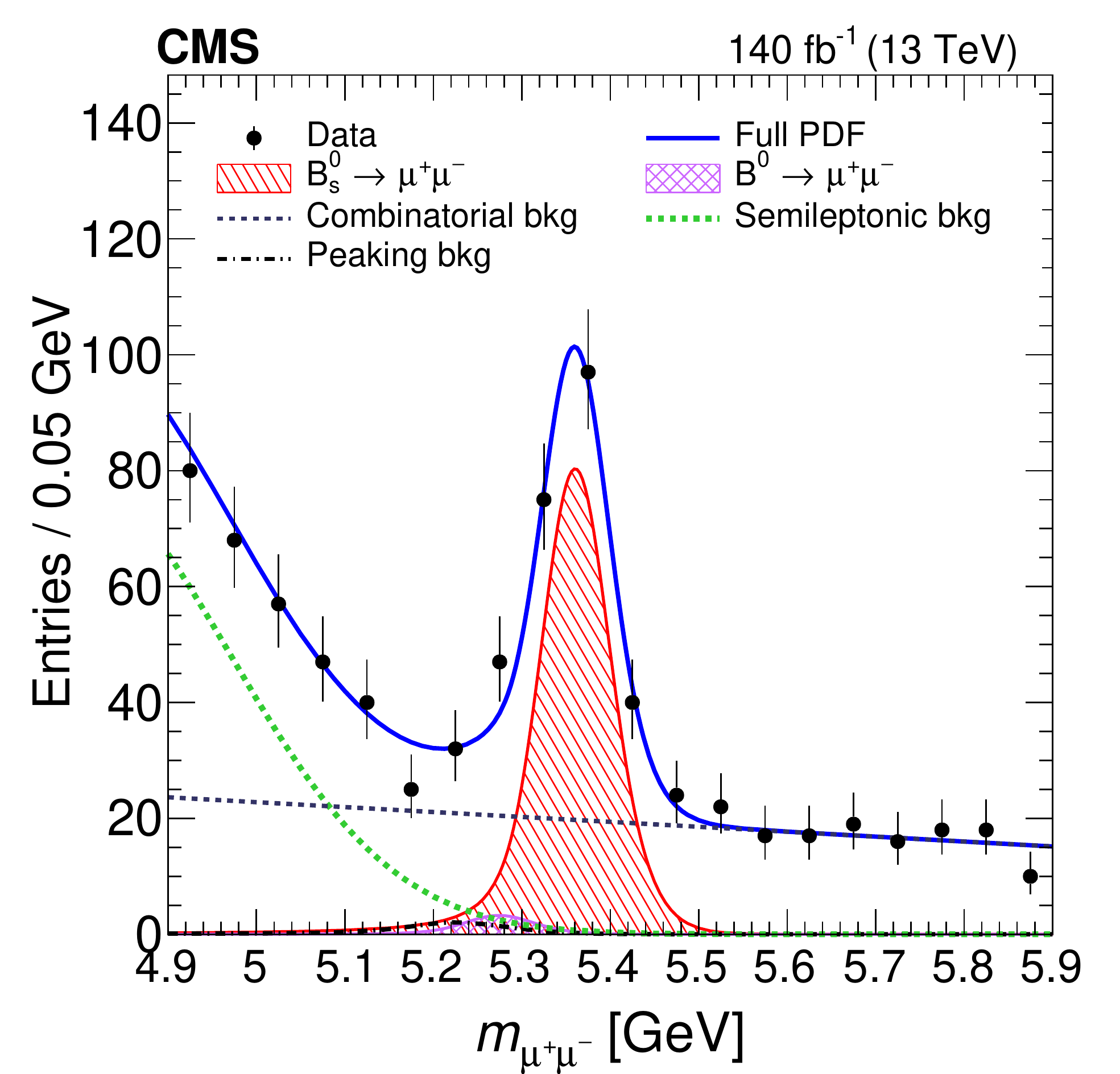}
    \includegraphics[width=0.48\textwidth]{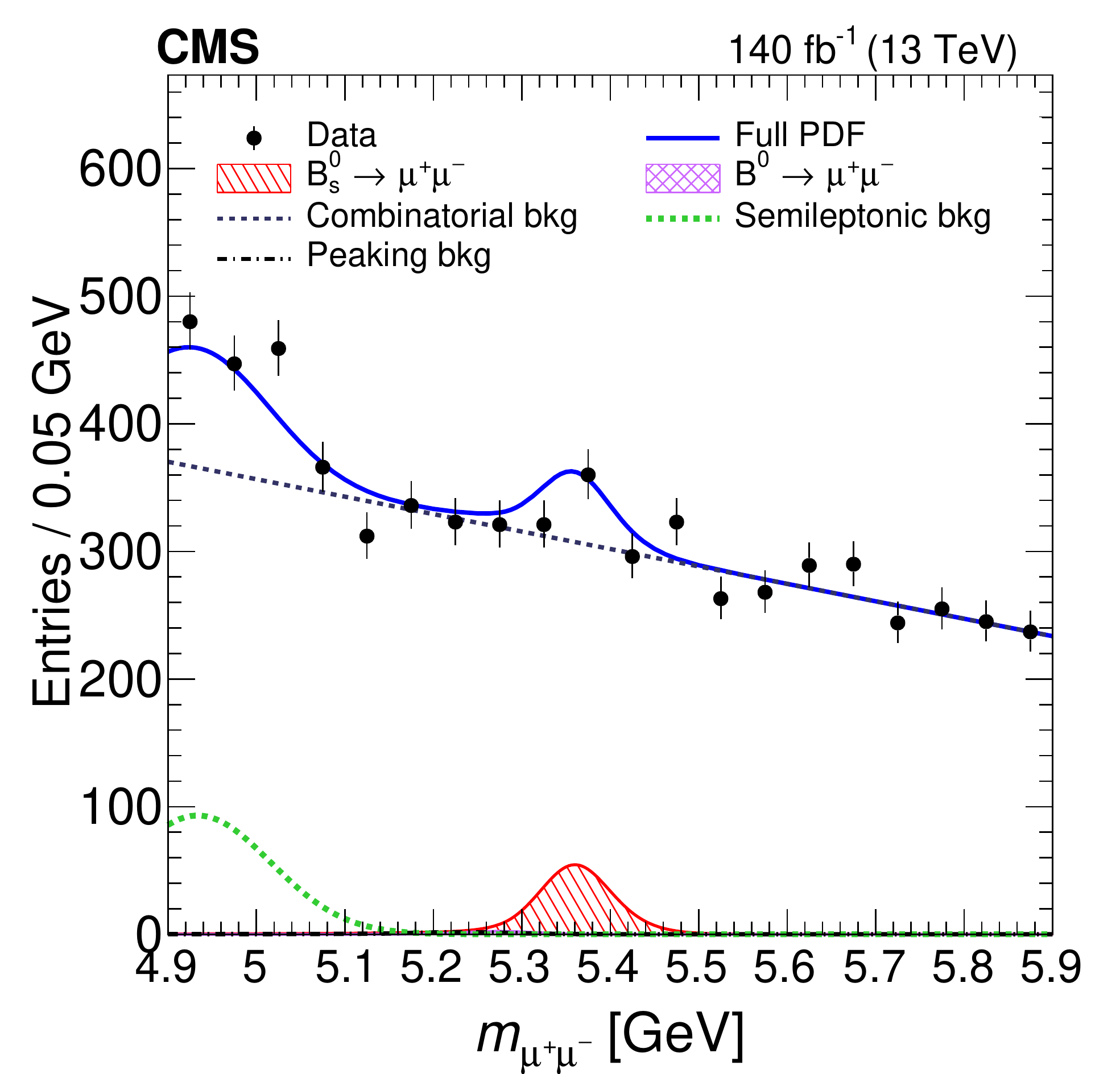}
    \caption{The projections on the dimuon mass axis of the fit to the branching fraction
      for the $\mvab>0.99$ (\cmsLeft) and $0.99>\mvab>0.90$ (\cmsRight) categories.
      The solid blue curves represent the corresponding projections of the
      final fit model, while the individual components of the fit are represented by the 
      dashed curves (backgrounds) and hatched histograms (signals).}
    \label{fig:projection_merge_unblind}
\end{figure}

\begin{figure}[tp]
\centering
    \includegraphics[width=0.48\textwidth]{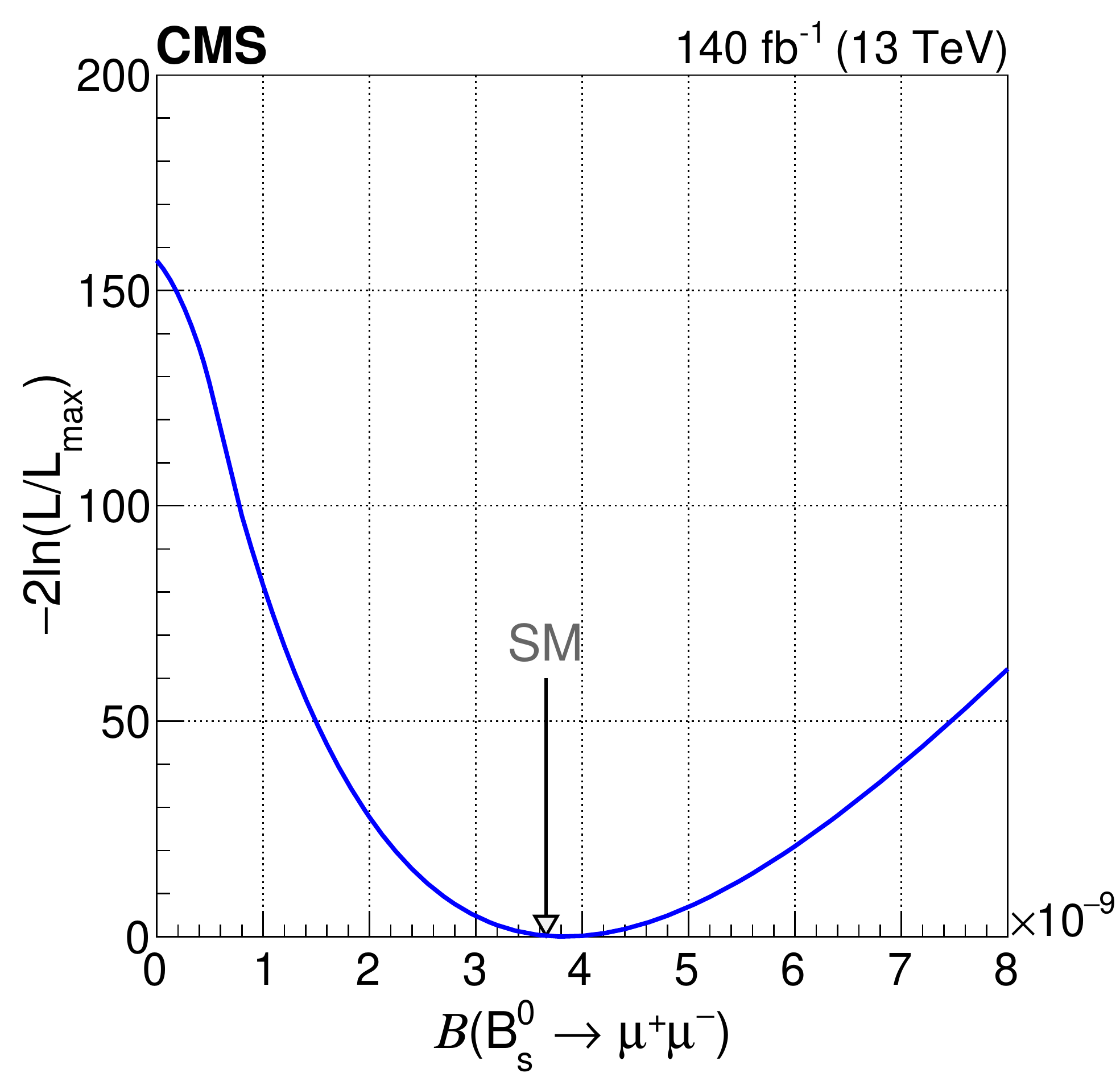}
    \includegraphics[width=0.48\textwidth]{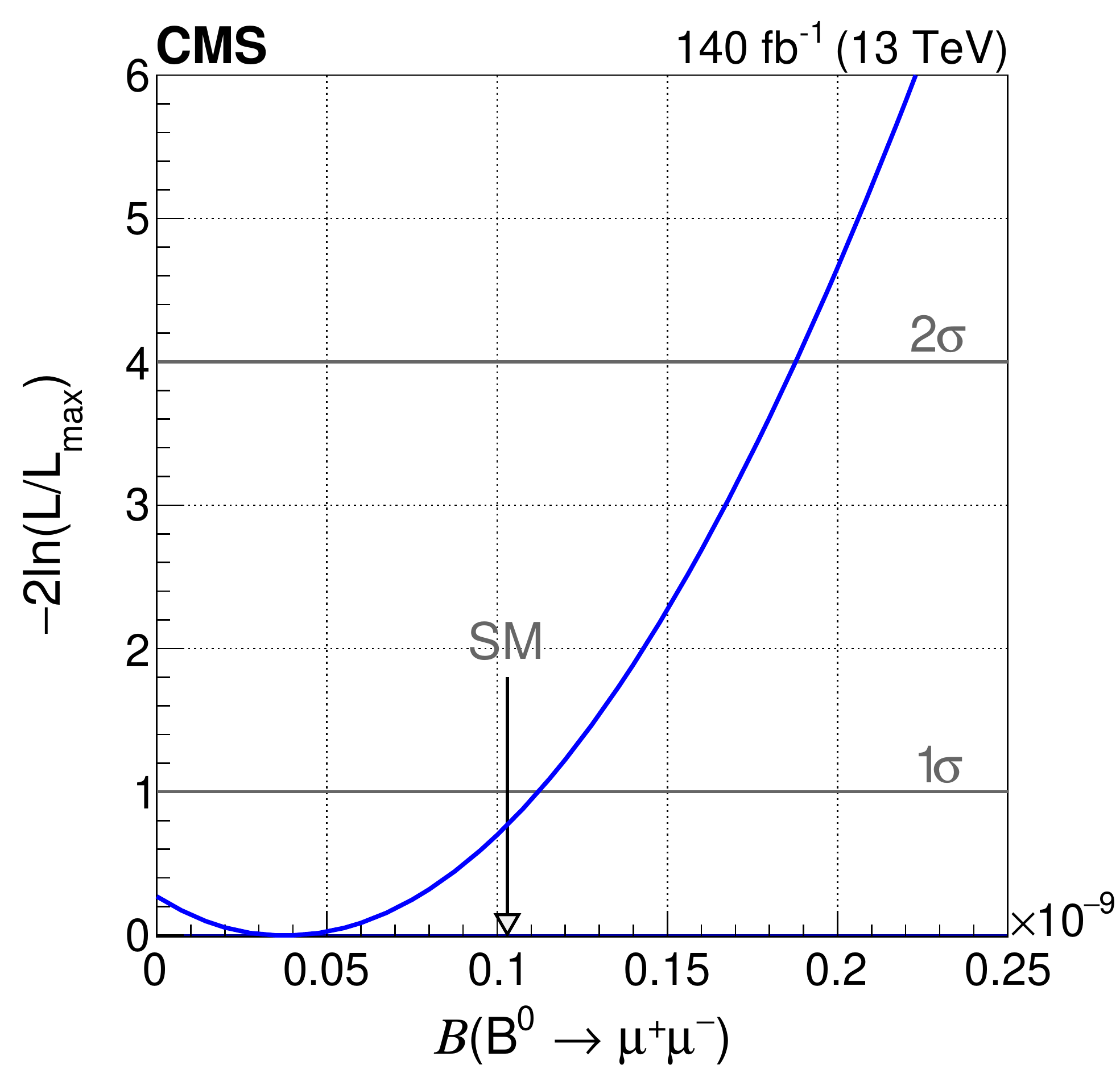}
    \includegraphics[width=0.48\textwidth]{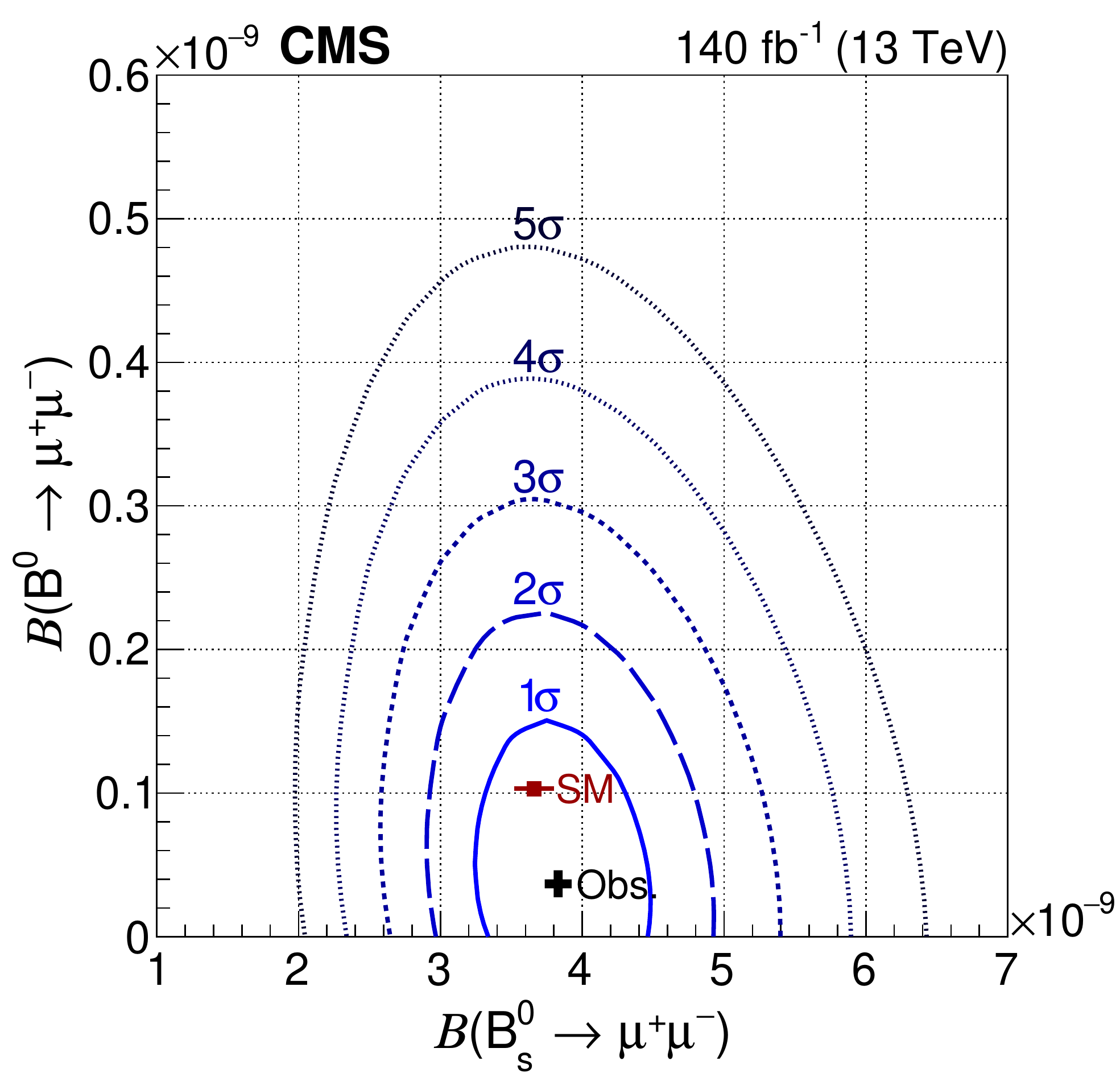}
    \caption{The profile likelihood as a function of \bsmm (\cmsLeft) and \bdmm
      (\cmsTopRight) decay branching fractions in 1D (\cmsTwoUpper) and in 2D (lower
      plot).  The contours in 2D enclose the regions with 1--5\unit{$\sigma$}
      coverage, where 1, 2, and 3\unit{$\sigma$} regions correspond to
      68.3, 95.4, and 99.7\% confidence levels, respectively.}
    \label{fig:nll_scans}
\end{figure}

The mass projections of the likelihood fits with all four data-taking periods combined together are shown
in Fig.~\ref{fig:projection_merge_unblind}. The event yields for each component of the
fit are summarized in Table~\ref{tab:unblind_yields_table}. The profile likelihood as a
function of the \bsmm and \bdmm branching fractions for 1D and 2D cases are
shown in Fig.~\ref{fig:nll_scans}. 

We also estimate the branching fractions using the \bjpsiphi decays for the
normalization. While this result is free from the explicit systematic uncertainty in the \fsfu
ratio, it depends on the \bjpsiphi branching fraction. At the moment, this
branching fraction measurement uses the \fsfu ratio measurement as an input,
but this dependence may be eliminated when new independent
measurements of the \bjpsiphi branching fraction become available, such as the
measurement planned by the Belle~II Collaboration
at the KEKB $\EE$ collider~\cite{Belle-II:2018jsg} using the $\PGUP{\mathrm{5S}}$ data.
Experimentally, the measurement based on the \bjpsiphi normalization channel
has slightly larger systematic uncertainties due to the presence of the second kaon in
the final state.

Taking the world average value of $\cbf(\bjpsiphi ) = (1.04\pm0.040) \times 10^{-3}$~\cite{Workman:2022ynf}, 
dominated by the LHCb measurement~\cite{LHCb:2021qbv},
and using the result of the \bjpsiphi normalization fit with
Eq.~\eqref{eq:bf_bs_jpsiphi}, we get:
\begin{linenomath}
\ifthenelse{\boolean{cms@external}}
{ 
\begin{multline*}
  \cbf(\bsmm) = \\
  = \left[ 4.02^{+0.40}_{-0.38}\stat\,^{+0.28}_{-0.23}\syst\,^{+0.18}_{-0.15}\,(\cbf)\right] \times 10^{-9},
\end{multline*}
} 
{ 
\begin{align*}
  \cbf(\bsmm) 
  = \left[ 4.02^{+0.40}_{-0.38}\stat\,^{+0.28}_{-0.23}\syst\,^{+0.18}_{-0.15}\,(\cbf)\right] \times 10^{-9},
\end{align*}
} 
\end{linenomath}
where the last uncertainty comes from the uncertainty in the \bjpsiphi branching fraction.

\begin{figure}[thp]
  \centering
  \includegraphics[width=\largePlotWidth]{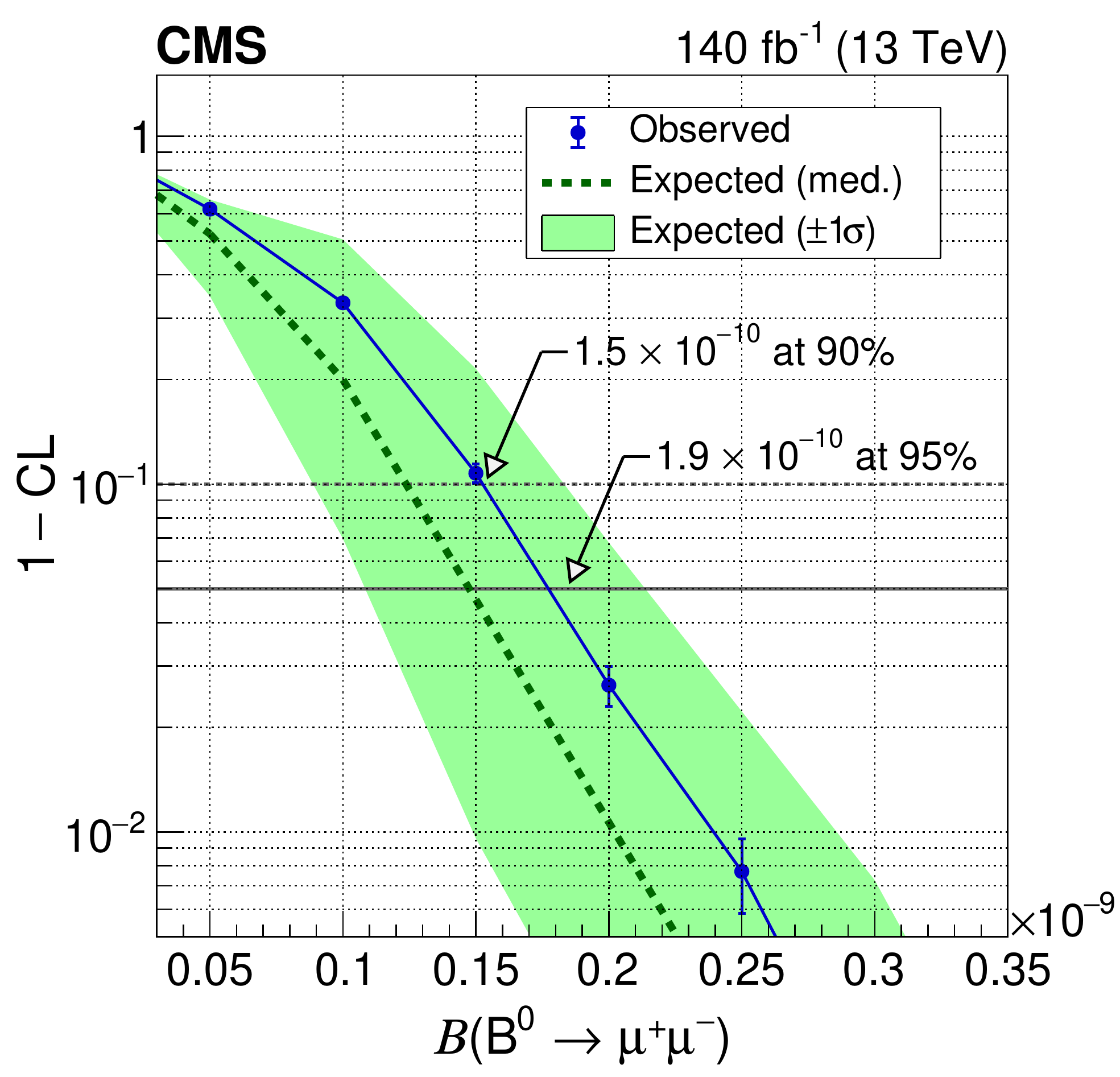}
  \caption{The upper limits on the \bdmm decay branching fraction using the \CLs method. The dashed line represents the
    expected median value of the quantity $1-\CL$ for the background-only hypothesis, while the solid line shows the observed value.
    The shaded region indicates the $\pm 1\sigma$ band.
  }
  \label{fig:bdstat_cl_limit}
\end{figure}

The 90 and 95\% confidence level (\CL) upper limits on \cbf(\bdmm) are evaluated using the \CLs
criterion~\cite{Junk,Read} and found to be
\begin{linenomath}
\begin{align*}
\cbf(\bdmm) &< 1.5 \times 10^{-10}~\text{at~90\%~\CL},\\
\cbf(\bdmm) &< 1.9 \times 10^{-10}~\text{at~95\%~\CL},
\end{align*}
\end{linenomath}
as shown in Fig.~\ref{fig:bdstat_cl_limit}.

\begin{figure}[thbp!]
\centering
    \includegraphics[width=\largePlotWidth]{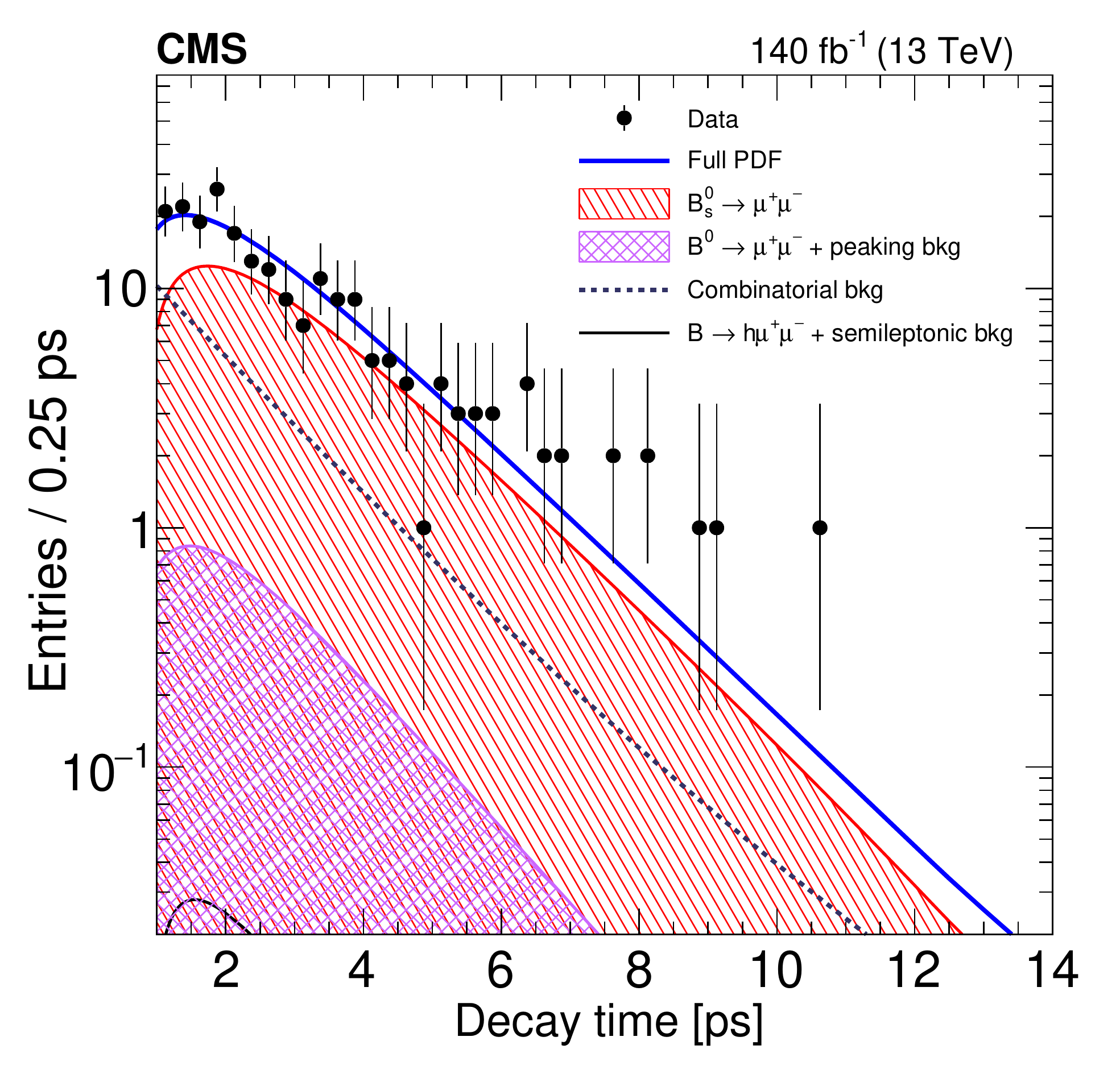}
    \caption{The UML fit projection on the decay time axis for the signal region $5.28 < m_{\MM} < 5.48\GeV$.}
    \label{fig:projection_decaytimemerge_signalonly}
\end{figure}

The effective lifetime for the \bsmm decay is found to be
\begin{linenomath}
\begin{align*}
\tau &= 1.83\,^{+0.23}_{-0.20}\stat\,^{+0.04}_{-0.04}\syst\unit{ps}.
\end{align*}
\end{linenomath}
The UML fit projection on the decay time axis for the signal region $5.28 < m_{\MM} < 5.48\GeV$ is shown in
Fig.~\ref{fig:projection_decaytimemerge_signalonly}. The observed lifetime is
consistent with the world average value of $1.624 \pm 0.009\unit{ps}$~\cite{Workman:2022ynf} within 1\unit{$\sigma$}, and therefore we do not correct
the corresponding selection efficiency when performing the branching fraction measurement.

\begin{table*}[hptb!]
  \centering
  \topcaption{The expected event yields for \bsmm ($N(\Bs)$), \bdmm ($N(\Bd)$), the combinatorial
    background ($N(\text{comb})$), the peaking background ($N(\text{peak})$), and the semileptonic background ($N(\text{semi})$)
    are summarized for each category (post-fit). The total expected and observed event yields are given in $N(\text{total})$ and Data
    column, respectively. Regions 0 and 1 refer to the ranges of 0.0--0.7 and 0.7--1.4, respectively, for the $\abs{\eta}$ of the most forward muon. The uncertainties are statistical only.}
    \cmsTable{
    \begin{tabular}{lcrrrrrrr}
      Data set & Region & $N(\Bs)$ & $N(\Bd)$ & $N(\text{comb})$ & $N(\text{peak})$ & $N(\text{semi})$ & $N(\text{total})$ & Data\\
      \hline
      \multicolumn{9}{c}{$\mvab>0.99$}\\
      \hline
      2016a & 0 & $5.3\pm 0.6$  & $0.2\pm 0.4$ &  $2.8\pm 1.9$     & $0.2\pm 0.1$ & $5.9\pm 1.2$  & $14.5\pm 2.3$   &  16  \\
      2016a & 1 & $9.4\pm 1.0$  & $0.4\pm 0.7$ &  $16.3\pm 5.6$    & $0.4\pm 0.2$ & $9.8\pm 1.9$  & $36.2\pm 5.4$   &  35  \\
      2016b & 0 & $6.3\pm 0.7$  & $0.3\pm 0.5$ &  $1.8\pm 2.0$     & $0.2\pm 0.1$ & $7.8\pm 1.5$  & $16.3\pm 2.8$   &  12  \\
      2016b & 1 & $9.9\pm 1.1$  & $0.4\pm 0.8$ &  $8.7\pm 5.6$     & $0.4\pm 0.2$ & $13.1\pm 2.5$ & $32.5\pm 5.1$   &  32  \\
      2017  & 0 & $23.5\pm 2.5$ & $1.0\pm 1.9$ &  $51.7\pm 11.0$   & $0.7\pm 0.3$ & $29.3\pm 4.9$ & $106.2\pm 9.8$  &  114 \\
      2017  & 1 & $33.8\pm 3.5$ & $1.4\pm 2.7$ &  $90.1\pm 14.0$   & $1.4\pm 0.5$ & $43.5\pm 6.9$ & $170.1\pm 12.3$ &  165 \\
      2018  & 0 & $34.4\pm 3.6$ & $1.4\pm 2.8$ &  $65.1\pm 12.5$   & $1.3\pm 0.5$ & $38.1\pm 6.0$ & $140.2\pm 11.1$ &  143 \\
      2018  & 1 & $49.9\pm 5.2$ & $2.0\pm 4.0$ &  $151.6\pm 17.5$  & $2.4\pm 1.0$ & $50.5\pm 7.8$ & $256.4\pm 15.4$ &  252 \\
      \hline
      \multicolumn{9}{c}{$0.99>\mvab>0.90$}\\
      \hline
      2016a & 0 & $4.8\pm 0.5$  & $0.2\pm 0.4$ & $118.1\pm 11.6$  & $0.2\pm 0.1$ & $8.3\pm 1.8$   & $131.6\pm 11.4$  & 132  \\
      2016a & 1 & $8.9\pm 1.0$  & $0.4\pm 0.7$ & $325.1\pm 19.1$  & $0.4\pm 0.2$ & $16.3\pm 3.3$  & $351.0\pm 18.7$  & 352  \\
      2016b & 0 & $5.5\pm 0.6$  & $0.2\pm 0.5$ & $107.9\pm 11.4$  & $0.2\pm 0.1$ & $10.7\pm 2.2$  & $124.5\pm 11.1$  & 126  \\
      2016b & 1 & $9.2\pm 1.0$  & $0.4\pm 0.7$ & $257.4\pm 17.3$  & $0.4\pm 0.2$ & $18.0\pm 3.5$  & $285.4\pm 16.9$  & 287  \\
      2017  & 0 & $15.1\pm 1.8$ & $0.6\pm 1.2$ & $638.2\pm 26.6$  & $0.7\pm 0.3$ & $26.1\pm 4.7$  & $680.7\pm 26.1$  & 683  \\
      2017  & 1 & $21.6\pm 2.5$ & $0.9\pm 1.8$ & $1431.2\pm 39.6$ & $1.0\pm 0.4$ & $43.7\pm 7.8$  & $1498.5\pm 38.7$ & 1498 \\
      2018  & 0 & $23.2\pm 2.7$ & $1.0\pm 1.9$ & $937.3\pm 33.6$  & $1.1\pm 0.5$ & $51.5\pm 9.9$  & $1014.2\pm 31.8$ & 1017 \\
      2018  & 1 & $34.1\pm 4.0$ & $1.4\pm 2.8$ & $2223.5\pm 50.6$ & $1.8\pm 0.7$ & $78.7\pm 14.0$ & $2339.5\pm 48.3$ & 2340 \\
    \end{tabular}
    }
  \label{tab:unblind_yields_table}
\end{table*}

\section{Summary}
\label{sec:summary}

Measurements of the branching fraction (\cbf) of the \bsmm decay and the effective \Bs meson lifetime
in this decay based on a data set of proton-proton collisions
at $\sqrt{s} = 13\TeV$ corresponding to an integrated luminosity of 140\fbinv have been presented and
found to be:
\begin{linenomath}
\ifthenelse{\boolean{cms@external}}
{ 
\begin{multline*}
  \cbf(\bsmm) = \\
  = \left[ 3.83^{+0.38}_{-0.36}~\stat\,^{+0.19}_{-0.16}\syst\,^{+0.14}_{-0.13}\,\text{(\fsfu)} \right] \times 10^{-9},
\end{multline*}
\begin{equation*}
\tau = 1.83\,^{+0.23}_{-0.20}\stat\,^{+0.04}_{-0.04}\syst\unit{ps}.
\end{equation*}
} 
{ 
\begin{align*}
  \cbf(\bsmm) 
  &= \left[ 3.83^{+0.38}_{-0.36}~\stat\,^{+0.19}_{-0.16}\syst\,^{+0.14}_{-0.13}\,\text{(\fsfu)} \right] \times 10^{-9},\\
\tau &= 1.83\,^{+0.23}_{-0.20}\stat\,^{+0.04}_{-0.04}\syst\unit{ps}.
\end{align*}
} 
\end{linenomath}
Both measurements are the most precise single measurements to date and
consistent with the standard model (SM) predictions and previous measurements
within one standard deviation. The relative total uncertainty in $\cbf$~is
reduced from 23 to 11\% compared with the previous CMS
measurement~\cite{CMS:2019bbr}, based on 2011-2012 and partial 13\TeV data sets,
while the central value is found to be somewhat higher. The 2016 data sample has
been re-analyzed and the new results supersede the ones from
Ref.~\cite{CMS:2019bbr}. The new analysis applied to the 2016 data yields a
central value similar to the original measurement, indicating that the shift in
the central value is driven mostly by the new data.

The search for the \bdmm decay has not revealed a significant event excess with
respect to the dominant combinatorial background prediction. The 95\% confidence level
upper limit on the branching fraction is found to be
\begin{linenomath}
\begin{align*}
\cbf(\bdmm) &< 1.9 \times 10^{-10}~\text{at~95\%~\CL}.
\end{align*}
\end{linenomath}
More data will be required to establish its existence and compare the result
with the SM predictions.

Compared with the latest LHCb measurement~\cite{LHCb:2021awg,LHCb:2021awg2}
\begin{linenomath}
\begin{align*}
\cbf(\bsmm) = (3.09^{+0.46}_{-0.43}\stat^{+0.15}_{-0.11}\syst)\times 10^{-9},
\end{align*}
\end{linenomath}
our result with the combined systematic uncertainty including the external uncertainties,
is about 1.2 standard deviations higher. These two measurements will shift the world average from
its current value of $\cbf(\bsmm) = (2.69^{+0.37}_{-0.35})\times
10^{-9}$~\cite{CMS:2020rox} to a larger value, more consistent with the SM
prediction, thus reducing the overall tension. The new measurement of the \bsmm
branching fraction is an important input to the global fits to the flavor data
(e.g., Ref.~\cite{Altmannshofer:2021qrr}) in light of the reported \bsll
anomalies (where lepton $\ell = \Pe$ or \Pgm).

The uncertainties in the branching fraction and effective lifetime measurements are dominated by the statistical
component, which means that significant improvements can be expected in the precision
of future measurements with the LHC Run 3 data.

The effective \Bs meson lifetime measurement in the \bsmm decay has achieved a precision
comparable with the lifetime difference between the heavy and light \Bs
meson mass eigenstates, thus offering sensitivity to potential beyond-the-SM physics effects
in the effective lifetime.

\begin{acknowledgments}
We congratulate our colleagues in the CERN accelerator departments for the excellent performance of the LHC and thank the technical and administrative staffs at CERN and at other CMS institutes for their contributions to the success of the CMS effort. In addition, we gratefully acknowledge the computing centres and personnel of the Worldwide LHC Computing Grid and other centres for delivering so effectively the computing infrastructure essential to our analyses. Finally, we acknowledge the enduring support for the construction and operation of the LHC, the CMS detector, and the supporting computing infrastructure provided by the following funding agencies: BMBWF and FWF (Austria); FNRS and FWO (Belgium); CNPq, CAPES, FAPERJ, FAPERGS, and FAPESP (Brazil); MES and BNSF (Bulgaria); CERN; CAS, MoST, and NSFC (China); MINCIENCIAS (Colombia); MSES and CSF (Croatia); RIF (Cyprus); SENESCYT (Ecuador); MoER, ERC PUT and ERDF (Estonia); Academy of Finland, MEC, and HIP (Finland); CEA and CNRS/IN2P3 (France); BMBF, DFG, and HGF (Germany); GSRI (Greece); NKFIH (Hungary); DAE and DST (India); IPM (Iran); SFI (Ireland); INFN (Italy); MSIP and NRF (Republic of Korea); MES (Latvia); LAS (Lithuania); MOE and UM (Malaysia); BUAP, CINVESTAV, CONACYT, LNS, SEP, and UASLP-FAI (Mexico); MOS (Montenegro); MBIE (New Zealand); PAEC (Pakistan); MES and NSC (Poland); FCT (Portugal); MESTD (Serbia); MCIN/AEI and PCTI (Spain); MOSTR (Sri Lanka); Swiss Funding Agencies (Switzerland); MST (Taipei); MHESI and NSTDA (Thailand); TUBITAK and TENMAK (Turkey); NASU (Ukraine); STFC (United Kingdom); DOE and NSF (USA).

\hyphenation{Rachada-pisek} Individuals have received support from the Marie-Curie programme and the European Research Council and Horizon 2020 Grant, contract Nos.\ 675440, 724704, 752730, 758316, 765710, 824093, 884104, and COST Action CA16108 (European Union); the Leventis Foundation; the Alfred P.\ Sloan Foundation; the Alexander von Humboldt Foundation; the Belgian Federal Science Policy Office; the Fonds pour la Formation \`a la Recherche dans l'Industrie et dans l'Agriculture (FRIA-Belgium); the Agentschap voor Innovatie door Wetenschap en Technologie (IWT-Belgium); the F.R.S.-FNRS and FWO (Belgium) under the ``Excellence of Science -- EOS" -- be.h project n.\ 30820817; the Beijing Municipal Science \& Technology Commission, No. Z191100007219010; the Ministry of Education, Youth and Sports (MEYS) of the Czech Republic; the Hellenic Foundation for Research and Innovation (HFRI), Project Number 2288 (Greece); the Deutsche Forschungsgemeinschaft (DFG), under Germany's Excellence Strategy -- EXC 2121 ``Quantum Universe" -- 390833306, and under project number 400140256 - GRK2497; the Hungarian Academy of Sciences, the New National Excellence Program - \'UNKP, the NKFIH research grants K 124845, K 124850, K 128713, K 128786, K 129058, K 131991, K 133046, K 138136, K 143460, K 143477, 2020-2.2.1-ED-2021-00181, and TKP2021-NKTA-64 (Hungary); the Council of Science and Industrial Research, India; the Latvian Council of Science; the Ministry of Education and Science, project no. 2022/WK/14, and the National Science Center, contracts Opus 2021/41/B/ST2/01369 and 2021/43/B/ST2/01552 (Poland); the Funda\c{c}\~ao para a Ci\^encia e a Tecnologia, grant CEECIND/01334/2018 (Portugal); the National Priorities Research Program by Qatar National Research Fund; MCIN/AEI/10.13039/501100011033, ERDF ``a way of making Europe", and the Programa Estatal de Fomento de la Investigaci{\'o}n Cient{\'i}fica y T{\'e}cnica de Excelencia Mar\'{\i}a de Maeztu, grant MDM-2017-0765 and Programa Severo Ochoa del Principado de Asturias (Spain); the Chulalongkorn Academic into Its 2nd Century Project Advancement Project, and the National Science, Research and Innovation Fund via the Program Management Unit for Human Resources \& Institutional Development, Research and Innovation, grant B05F650021 (Thailand); the Kavli Foundation; the Nvidia Corporation; the SuperMicro Corporation; the Welch Foundation, contract C-1845; and the Weston Havens Foundation (USA).  
\end{acknowledgments}
\bibliography{auto_generated}
\cleardoublepage \appendix\section{The CMS Collaboration \label{app:collab}}\begin{sloppypar}\hyphenpenalty=5000\widowpenalty=500\clubpenalty=5000\input{BPH-21-006-public-authorlist.tex}\end{sloppypar}
\end{document}

%% file: BPH-21-006-public-authorlist.tex
\cmsinstitute{Yerevan Physics Institute, Yerevan, Armenia}
{\tolerance=6000
A.~Tumasyan\cmsAuthorMark{1}\cmsorcid{0009-0000-0684-6742}
\par}
\cmsinstitute{Institut f\"{u}r Hochenergiephysik, Vienna, Austria}
{\tolerance=6000
W.~Adam\cmsorcid{0000-0001-9099-4341}, J.W.~Andrejkovic, T.~Bergauer\cmsorcid{0000-0002-5786-0293}, S.~Chatterjee\cmsorcid{0000-0003-2660-0349}, K.~Damanakis\cmsorcid{0000-0001-5389-2872}, M.~Dragicevic\cmsorcid{0000-0003-1967-6783}, A.~Escalante~Del~Valle\cmsorcid{0000-0002-9702-6359}, P.S.~Hussain\cmsorcid{0000-0002-4825-5278}, M.~Jeitler\cmsAuthorMark{2}\cmsorcid{0000-0002-5141-9560}, N.~Krammer\cmsorcid{0000-0002-0548-0985}, L.~Lechner\cmsorcid{0000-0002-3065-1141}, D.~Liko\cmsorcid{0000-0002-3380-473X}, I.~Mikulec\cmsorcid{0000-0003-0385-2746}, P.~Paulitsch, F.M.~Pitters, J.~Schieck\cmsAuthorMark{2}\cmsorcid{0000-0002-1058-8093}, R.~Sch\"{o}fbeck\cmsorcid{0000-0002-2332-8784}, D.~Schwarz\cmsorcid{0000-0002-3821-7331}, M.~Sonawane\cmsorcid{0000-0003-0510-7010}, S.~Templ\cmsorcid{0000-0003-3137-5692}, W.~Waltenberger\cmsorcid{0000-0002-6215-7228}, C.-E.~Wulz\cmsAuthorMark{2}\cmsorcid{0000-0001-9226-5812}
% ...
\par}
\vskip\cmsinstskip
\dag:~Deceased\\
$^{1}$Also at Yerevan State University, Yerevan, Armenia\\
$^{2}$Also at TU Wien, Vienna, Austria\\
% ...
